%% file: eddyGSOpt2025.tex
\documentclass[twocolumn,letterpaper]{IEEEAerospaceCLS}  % only supports two-column, letterpaper format

\normalsize % Needs to be put in before \usepackage{booktabs} for it to work

\usepackage[]{graphicx}    % We use this package in this document
\usepackage[hidelinks]{hyperref}
\usepackage{cite}
\usepackage[detect-all,per-mode=symbol,binary-units=true]{siunitx}
\usepackage{longtable}     % For long tables
\usepackage{geometry}      % For adjusting page margins
\usepackage{array}		   % Provide aligned
\usepackage{amsmath,amstext,amssymb} % Provides significant math features
\usepackage{algorithm}
\usepackage[noend]{algpseudocode}
\usepackage{booktabs}      % For better table lines
\usepackage{cleveref} 	   % Provides \Cref
\usepackage{graphicx}
\usepackage{subcaption}

\newcommand{\ignore}[1]{}  % {} empty inside = %% comment

\begin{document}
\title{Optimal Ground Station Selection for \\ Low-Earth Orbiting Satellites}

\author{%
Duncan Eddy, Michelle Ho, and Mykel J. Kochenderfer\\ 
Stanford University\\
496 Lomita Mall\\
Stanford, CA 94305\\
deddy@stanford.edu, mtho@stanford.edu, mykel@stanford.edu
%%%% IMPORTANT: Use the correct copyright information--IEEE, Crown, or U.S. government. %%%%%
\thanks{\footnotesize 979-8-3503-5597-0/25/$\$31.00$ \copyright2025 IEEE}              % This creates the copyright info that is the correct 2025 data.
}

\maketitle

\thispagestyle{plain}
\pagestyle{plain}

\begin{abstract}
This paper presents a solution to the problem of optimal ground station selection for low-Earth orbiting (LEO) space missions that enables mission operators to precisely design their ground segment performance and costs. Space mission operators are increasingly turning to Ground-Station-as-a-Service (GSaaS) providers to supply the terrestrial communications segment to reduce costs and increase network size. However, this approach leads to a new challenge of selecting the optimal service providers and station locations for a given mission. We consider the problem of ground station selection as an optimization problem and present a general solution framework that allows mission designers to set their overall optimization objective and constrain key mission performance variables such as total data downlink, total mission cost, recurring operational cost, and maximum communications time-gap. We solve the problem using integer programming (IP). To address computational scaling challenges, we introduce a surrogate optimization approach where the optimal station selection is determined based on solving the problem over a reduced time domain. Two different IP formulations are evaluated using randomized selections of LEO satellites of varying constellation sizes. We consider the networks of the commercial GSaaS providers Atlas Space Operations, Amazon Web Services (AWS) Ground Station, Azure Orbital Ground Station, Kongsberg Satellite Services (KSAT), Leaf Space, and Viasat Real-Time Earth. We compare our results against standard operational practices of integrating with one or two primary ground station providers.
\end{abstract}

\tableofcontents

\input{sections/introduction} % Note: Need to use \input instead of \include because \include inserts a \clearpage prior to the inserted text which throws off formatting

\input{sections/problem_formulation}

\input{sections/experiments}

\input{sections/conclusions}

%%%%%%%%%%%%%%%%%%%%%%%%%%%%%%%%%%%%%%%%%%%%%%%%%%%%%%%%%%%%%%%%%%%%%%%%%%%%%%%%%%%%%%%%%%%%%%%%%
\appendices{}              % note there is no {} to put a title. Each appendix has its own title
%%%%%%%%%%%%%%%%%%%%%%%%%%%%%%%%%%%%%%%%%%%%%%%%%%%%%%%%%%%%%%%%%%%%%%%%%%%%%%%%%%%%%%%%%%%%%%%%%
% For a single appendix, use the \appendix{} keyword and do not use the \section command.
%%%%%%%%%%%%%%%%%%%%%%%%%%

The software utilized to formulate the contact optimization problem and prepare this publication is available at \href{https://github.com/sisl/ground-station-optimizer}{\textcolor{blue}{https://github.com/sisl/ground-station-optimizer}}.

\input{sections/appendix_gs_list}

\input{sections/appendix_contact_assumption}

\input{sections/appendix_min_cost_opt}

\input{sections/appendix_max_data_opt}

\input{sections/appendix_max_gap_opt}
\acknowledgments
The authors would like to thank Arthur Kvalheim Merlin at KSAT, Jai Dialani at Leaf Space, and Aaron Hawkins at Viasat for their help in providing and reviewing current ground station site lists for this work.

%%%%%%%%%%%%%%%%%%%%%%%%%%%%%%%%%%%%%%%%%%%%%%%%%%%%%%%%%%%%%%%%%%%%%%%%%%%%%%%%%%%%%%%%%%%%%%%%%%%%%%
\bibliographystyle{IEEEtran}
\bibliography{eddyGSOpt2025.bib}

%%%%%%%%%%%%%%%%%%%%%%%%%%%%%%%%%%%%%%%%%%%%%%%%%%%%%%%%%%%%%%%%%%%%%%%%%%%%%%%%%%%%%%%%%%%%%%%%%%%%%%
\thebiography
%% This biostyle allows you to insert your photo size 1in X 1.25in

\begin{biographywithpic}{Duncan Eddy}{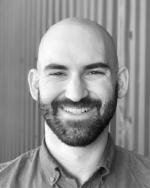}
is a postdoctoral research fellow in the Stanford Intelligent Systems Laboratory (SISL) and the Executive Director of the Center for AI Safety at Stanford University. His research focuses on decision-making in safety-critical, climate, and space systems. He received B.S. in Mechanical Engineer from Rice University in 2013, and PhD in Aerospace Engineering from Stanford University in 2021. Prior to returning to Stanford he was the Director of Space Operations at Capella Space Corporation where he built a fully automated constellation tasking and delivery system. He later founded and led the Constellation Management and Space Safety Organization at Amazon's Project Kuiper.
\end{biographywithpic} 

\begin{biographywithpic}{Michelle Ho}{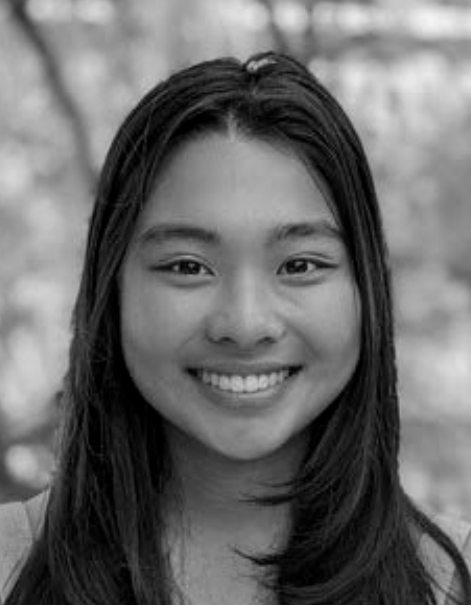}
is a PhD candidate in the Stanford Intelligent Systems Laboratory (SISL) in the department of Aeronautics and Astronautics at Stanford University. Her research focuses on decision making under uncertainty in complex environments, spacecraft autonomy, and optimal and learning-based control. She received her B.S. in Mechanical and Aerospace Engineering and a minor in Robotics and Intelligent Systems from Princeton University in 2023.
\end{biographywithpic}

\begin{biographywithpic}{Mykel Kochenderfer}{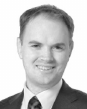}
 is an Associate Professor of Aeronautics and Astronautics and Associate Professor, by courtesy, of Computer Science at Stanford University. He is the director of the Stanford Intelligent Systems Laboratory (SISL), conducting research on advanced algorithms and analytical methods for the design of robust decision making systems. Prior to joining the faculty in 2013, he was at MIT Lincoln Laboratory where he worked on airspace modeling and aircraft collision avoidance. He received his Ph.D. from the University of Edinburgh in 2006 where he studied at the Institute of Perception, Action and Behaviour in the School of Informatics. He received B.S. and M.S. degrees in computer science from Stanford University in 2003.

\end{biographywithpic}

\end{document}

%% file: sections/introduction.tex
\section{Introduction}
\label{sec:introduction}

In recent years, easier access to space has led to a significant rise in the new constellations and missions being launched by commercial, government, and academic organizations. This change has been driven by new launch providers, cheaper ride-sharing launch opportunities, and the successful demonstration commercial-off-the-shelf-based small satellites. Historically, mission operators have provisioned dedicated ground segments to support each satellite or constellation separately. However, this leads to significant up-front capital expenditures, poor asset utilization rates, limited communications opportunities, and low data downlink volumes.

To address these challenges, the industry has moved increasingly to a GSaaS model where ground station providers build out a global network of station locations and  mission operators subsequently contract with providers to integrate with their network. This model reduces mission operator capital expenditure costs by spreading the cost over multiple customers while simultaneously increasing the usage of each antenna. Mission operators also benefit from increased scalability by being able to access existing global antenna networks, increasing the number of potential communications opportunities. This in turn improves data downlink volumes, reduces communication gaps, enables greater operational flexibility, and reduces mission lead-times.

\begin{figure*}[h!t]
\centering
\includegraphics[width=\textwidth]{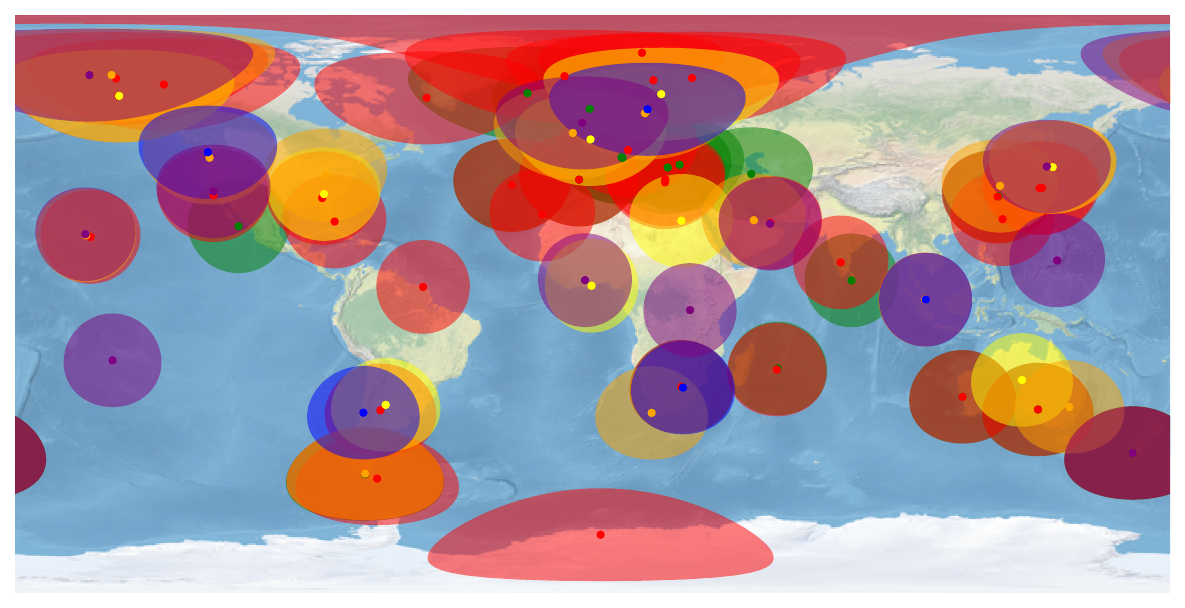}
\caption{Plot of all currently operational ground stations and associated communications cones for Atlas (purple), AWS (orange), Azure (blue), KSAT (red), Leaf Space (green), and Viasat (yellow). Communications cones assume \SI{525}{\km} altitude and $10^{\circ}$ minimum elevation angle. See Appendix \ref{sec:appendix_station_list} for station list.} 
\label{fig:all_ground_stations}
\end{figure*}

However, this new model also comes with its own challenges. There are integration and operational costs associated with using a provider. Additionally, the overall mission performance parameters for data downlink volume and contact frequency depend the combined selection of all station locations as well as the satellite orbits. Selecting the providers and stations to support a mission involves weighing considerations such as desired coverage, operational costs, latency and data throughput, uplink and downlink capabilities, reliability, and scheduling availability. This selection is complicated by the number of different potential providers, large selection of potential locations, and varying cost models between providers. Between the 6 largest ground station providers, there are over 91 ground station locations currently in operation. The providers and stations considered are listed in Appendix \ref{sec:appendix_station_list}. The global distribution and coverage of these stations is shown in \cref{fig:all_ground_stations}. This paper presents a solution to the ground station selection problem based on an integer programming (IP) formulation.

% Note: the hard-coded ref is bad, but otherwise with this template \Cref gives "Section A"

Early work proposed the idea federated ground station networks, a loose, virtual association of global stations to increase mission data downlink access while simultaneously lowering cost and reducing integration barriers~\cite{cutler2002federated,leveque2007global,white2015satnogs}. The primary focus of these works was on the software engineering challenges and architectural approaches for networking, scheduling, and data dissemination for a multi-mission, multi-operator ground station network.

In terms of station selection, past authors have focused on improving ground station placement for reducing downlink latency~\cite{vasisht2020distributed} and promoting site diversity to evade problematic weather conditions~\cite{fuchs2015ground,delportillo2017optimal,wrona2023intelligent}. Research in this area has focused on leveraging historical cloud data to optimize optical ground station network site selection, which is significantly impacted by adverse weather, to ensure high availability and minimize the number of required stations. Subsequent works account for variable cloud coverage using advanced models, historical data, and machine learning techniques~\cite{poulenard2015ground,net2016approximation,lyras2017cloud,lyras2018optimum,lyras2019optimizing}. Efrem et al.~consider ground station installation costs, but only optimize for outage probability independent of weather conditions~\cite{efrem2020minimizing,efrem2020globally}. Other works focus on ground station activity planning and scheduling through various optimization techniques, to maximize contact time and prioritize mission requirements~\cite{corrao2012ground,lala2015local,falone2018ground}, though these works do not consider the problem selecting the stations themselves and instead optimize mission plans over a pre-selected set of stations. 

% Know the direct "Efrem et al." is bad but \citeauthor appears to be broken with this document class

Similar problems to ground station placement include sensor placement~\cite{cardei2006energy}, facility/warehouse location placement~\cite{owen1998strategic}, antenna selection~\cite{gao2018massive}, computing resource allocation~\cite{wei2022multiobjective}, ambulance routing~\cite{tlili2017swarm}, and surveillance drone deployment~\cite{hu2019joint}. These problems often consider multiple objectives including coverage, network connectivity, detecting desired events or targets, monitoring environment factors, redundancy and fault tolerance, and placement for future, dynamically changing conditions. They have often been optimized with genetic algorithms~\cite{wang2009survey,huang2021genetic,shen2011pareto}.

Mission operators face the challenging decision problem of determining which ground station providers and ground stations to select to support their mission to optimize overall system performance. This paper presents a solution to the ground station selection problem that uses integer programming to solve a the surrogate optimization of selecting individual contact opportunities to maximize the selected objective within the bounds of system constraints. The providers and locations are then extracted based on the selected contact opportunities. The approach is applied to optimize the station selection across six current GSaaS providers\textemdash Atlas Space Operations, AWS Ground Station, Azure Orbital\footnote{Azure Orbital announced they plan to cease operations on December 18, 2024, after the initial submission of this work. Their stations have been retained in the manuscript since the core method optimizes over potential providers and is unaffected by their deprecation.}, KSAT, Leaf Space, and Viasat. We consider three different optimization objectives of maximizing total data downlink, minimizing total cost, and minimizing the maximum time gap between contacts. The viability of the approach is demonstrated by comparing the optimized station selection to a traditional fixed-provider solution in simulated scenarios where the provider-associated costs are randomized between trials. %We also present an example application of optimizing the ground station network for a commercial Earth observation constellation.

%% file: sections/problem_formulation.tex
\section{Problem Formulation}
\label{sec:problem_formulation}

We consider the problem of selecting a ground station network that optimizes a single performance objective while adhering to relevant selection constraints. The mission may include one or more satellites defined by set $\mathcal{S}$. The set of potential ground station providers $\mathcal{P}$ and station locations $\mathcal{L}$ are known and the problem is simply determining which locations should be selected. The mission start time $t_{opt}^s$ and end time $t_{opt}^e$ define the optimization window with total duration $T_{opt} = t_{opt}^e - t_{opt}^s$.

If we had an oracle that could compute all possible contact opportunities for all spacecraft over the entire mission duration using perfect knowledge of each object's predicted trajectory, then the ground station selection problem would be equivalent to the satellite task planning problem of selecting ground contact opportunities that optimize the desired objective. However, due to non-conservative orbital perturbations, it is not possible to accurately predict spacecraft orbits in LEO over a multi-year mission. Therefore, we instead consider the surrogate problem of optimizing the provider and location selection on a short segment of the mission duration defined by simulation start time $t_{sim}^s$ and end time $t_{sim}^e$ with total duration $T_{sim}$. The optimal station selection for this duration is considered to be the optimal selection over the entire mission.

This assumption is appropriate so long as the simulation window is long enough such that the distribution of contacts over the simulation period is representative of the distribution of contacts over the entire mission. Appendix~\ref{sec:appendix_prediction_assumption} discusses this in further depth, and finds that a propagation window of 7 to 10 days is sufficient, though at least 20 days is recommended. Alternatively if the mission uses repeat ground-track orbits and the simulation window matches the repetition period, then the optimization of providers and locations of the surrogate problem is exactly the same as optimizing over the entire mission duration. %The validity of this assumption and selection of the simulation sub-window is discussed further in Appendix \ref{sec:appendix_prediction_assumption}.

The problem of provider and location selection becomes the problem of optimizing selected contact opportunities for a set of satellites\textemdash the optimal providers and locations are simply the providers and locations corresponding to the selected contacts in the planning problem. There has been extensive work in the domain of satellite task planning with many different algorithmic approaches presented in literature since it was first presented by Hall et al.~\cite{hall1994maximizing}. Solution approaches include dynamic programming~\cite{augenstein2014optimal}, ant-colony optimization~\cite{iacopino2013novel}, mixed-integer linear programming~\cite{augenstein2016optimal}, Monte Carlo tree search~\cite{eddy2020markov}, and maximum independent set local search heuristics~\cite{eddy2021maximum}.

For this paper, we adopt an integer programming approach inspired by past mixed-integer linear programming work due to the unique benefits of this approach. If an optimization problem can be expressed as an IP, it can then be solved using software libraries such as Gurobi \cite{gurobi} or COIN-OR \cite{lougee2003common}. These libraries are fast, efficient, and have undergone extensive validation and use across numerous disciplines. Most importantly, as part of the solution to any problem, the solver returns an optimality certificate that states whether the returned solution is globally optimal, is suboptimal, or if the problem is infeasible and there is no possible valid solution. This certificate is useful for a system engineering in understanding the overall performance of the mission ground segment. The next step is to express the ground station optimization problem as an IP.

\subsection{IP Formulation}

%A general formulation of an IP is
%\begin{equation}
%\begin{aligned}
%    &\underset{\mathbf{x}}{\text{minimize}} && \mathbf{c}^\top \mathbf{x} \\
%    &\text{subject to} && \mathbf{A} \mathbf{x} \leq \mathbf{b}, \\
%    & && \mathbf{C} \mathbf{x} = \mathbf{d}, \\
%    & && \mathbf{x} \geq 0, \\
%    & && \mathbf{x} \in \mathbb{Z}^{\textit{n}} \\
%\end{aligned}
%\label{eqn:ip_general}
%\end{equation}
%The objective is to find the values of the design variables defined by vector \textbf{x} that maximize the linear sum weighted by vector \textbf{c}. The design variables are constrained by a set on inequality constraints defined by matrix weights \textbf{A} and vector limits \textbf{b} and set of equality constraints defined by matrix weights \textbf{C} and vector values \textbf{d}. Finally, all design variables are constrained by be positive integers in the set of $n$-dimensional integer vectors $\mathbb{Z}^{\textit{n}}$.

An integer program is a variation of a linear program where the objective and constraints are linear functions and all decision variables are integers. To express the ground station optimization problem as an IP, we consider a set of ground station providers $\mathcal{P}$. For each provider, there is a tuple $(p, P) \in \mathcal{P}$, where $p \in \{0,1\}$ is a design variable that is 1 if the provider is selected and 0, otherwise, and $P$ is a data-object that provides constant-information associated with that specific provider. Each provider has a set of ground station locations denoted ${\mathcal{L}}^{P}$, while the set of all stations is denoted $\mathcal{L}$. Each station $(l, L) \in {\mathcal{L}}$ has an associated design variable $l \in \{0,1\}$. The location $L$ has a fixed data rate $L_{dr}$. The ground network is optimized with respect to individual satellites $S \in \mathcal{S}$. Each satellite has a fixed data rate $S_{dr}$.

From $\mathcal{P}$, $\mathcal{L}$, and $\mathcal{S}$, we compute the set of all contacts $\mathcal{C}$ over the simulation window defined by $(t_{sim}^s, t_{sim}^e)$. For each contact $(c, C) \in \mathcal{C}$ there is a binary decision variable $c \in \{0,1\}$ that indicates whether contact $C$ has been selected. There are a number of constant variables associated with an individual contact $C$, such as the contact start time $C_{start}$, contact end time $C_{end}$. $C_{duration}$ is the total duration of a contact. The contact data rate $C_{dr}$ is 
\begin{equation}
\begin{aligned}
	C_{dr} = \min(L_{dr},S_{dr})
\end{aligned}
\label{eqn:contact_datarate}
\end{equation}
The location associated with a particular contact is denoted $C^L$ and the satellite associated with a contact is denoted $C^S$. We formulate different constraints by considering different subsets contacts associated with a particular provider ${\mathcal{C}}^P \in {\mathcal{C}}$, station ${\mathcal{C}}^L \in {\mathcal{C}}$, or satellite ${\mathcal{C}}^S \in {\mathcal{C}}$. Finally, we introduced auxiliary decision variable $v^{S,L} \in \{0,1\}$ to indicate whether satellite $S$ uses location $L$.

We also introduce a number of constants of the problem. $e^P_{integ}$ is the one-time expense associated with the engineering work to integrate with provider $P$. $e^{L}_{setup}$ is the cost to setup location $L$, which includes any one-time costs such as procuring radios, servers, or other equipment to support operations at the location. $e^{L}_{monthly}$ is the monthly recurring cost associated with using location $L$. This cost includes, but is not limited to, any costs associated with internet, power, security charged when the location is used. It is incurred if even only one contact is taken at that location. $e^{L}_{license}$ is the one-time cost associated with licensing a satellite to communicate with a specific station. $e^{L}_{pass}$ is the fixed cost of taking a contact at location $L$, while $e^{L}_{minute}$ is the cost-per-minute associated with using an antenna. Providers implement either a fixed-cost-per-pass or cost-per-minute pricing, but not both.

\subsection{Objective Functions}

There are three potential objective functions which a mission designer might want to consider for the optimization problem depending on the primary systems engineering goal for the mission.

The \textit{minimum cost objective} represents the goal of minimizing the total cost of the mission above all other concerns. This may be appropriate for budget-constrained missions. The total cost downlinked over the entire mission is
\begin{equation}
\begin{aligned}
    \sum_{(p,P) \in \mathcal{P}}e^P_{integ}p \, + \, \sum_{P \in \mathcal{P}}\sum_{(l,L) \in {\mathcal{L}}^P}e^L_{setup}l \, + \\
    \, \frac{12 \times T_{opt}}{365.25 \times 86400 \times T_{sim}}\sum_{P\in \mathcal{P}}\sum_{(l,L) \in {\mathcal{L}}^P}e^L_{monthly}l \, + \\
    \, \sum_{S \in \mathcal{S}}\sum_{P \in \mathcal{P}}\sum_{L \in {\mathcal{L}}^P} e^{L}_{license}v^{L,S} \, + \\
    \, \frac{T_{opt}}{T_{sim}}\sum_{P \in {\mathcal{P}}}\sum_{L \in {\mathcal{L}}^P}\sum_{(c,C) \in {\mathcal{C}}^{L}}(e^{L}_{minute}C_{duration} + e^L_{pass})c
\end{aligned}
\label{eqn:obj_min_cost}
\end{equation}
The constant weighting factors in front of the terms for the monthly and contact costs ensures that the objective function represents the total cost over the entire mission duration.

The \textit{maximum data downlink objective} represents the desire to maximize the total amount of data transmitted over the mission duration. This objective is appropriate for missions that are expected to be data-constrained and designing a station network that is able to maximize the total data return for the mission is the primary concern. The total data downlinked over the entire mission is
\begin{equation}
\begin{aligned}
	\frac{T_{opt}}{T_{sim}}\sum_{P \in {\mathcal{P}}}\sum_{L \in {\mathcal{L}}^P}\sum_{(c,C) \in {\mathcal{C}}^L}C_{dr}C_{duration}c
\end{aligned}
\label{eqn:obj_max_data}
\end{equation}
The objective is weighted by $\frac{T_{opt}}{T_{sim}}$ so that the optimal value of objective is the total data volume able to be downlinked over the mission.

The last objective is the \textit{minimum max-gap objective}, which seeks to minimize the maximum time gap between contacts for each satellite. This objective reflects a desire to design an operationally responsive ground network that ensures all vehicles are in regular contact with the ground. For an Earth observation mission this objective would bound the maximum latency of uplinking new tasking requests or downlinking data after collection. To implement this constraint we introduce an auxiliary binary decision variable $y_{ij}$ that is 1 if contact $j$ is the next contact after contact $i$ to be taken for a given satellite, $s$. The optimization problem then is
\begin{subequations}
\begin{align}
    &\underset{\mathbf{x}}{\text{minimize}} && G_{max} \label{eqn:obj_min_max_gap_1} \\ 
    &\text{subject to} && \sum_{j>i}y_{ij} = c_i, \forall \; (c,C) \in {\mathcal{C}}^S, S \in {\mathcal{S}}, \label{eqn:obj_min_max_gap_2} \\ % \forall i \in |X^{p,l}|
    & && \hspace{2em} i,j \in \{0,\ldots,|{\mathcal{C}}^S|\}\; \text{s.t.} \\
    & && \; C_{start,j} > C_{start,i} \\
    & && (C_{start,j} - C_{end,i})y_{ij} \leq G_{max},  \label{eqn:obj_min_max_gap_3}\\
    & && y_{ij} \leq c_i, \label{eqn:obj_min_max_gap_4}\\
    & && y_{ij} \leq c_j, \label{eqn:obj_min_max_gap_5}\\
    & && y_{ij} \in \{0,1\}, \label{eqn:obj_min_max_gap_6}\\
    & && c \in \{0,1\},  \label{eqn:obj_min_max_gap_7}
\end{align}
\label{eqn:obj_min_max_gap}
\end{subequations}

% Know the \eqref is not ideal and Cref is preferred, but this template doesn't play nicely with cleverref

Here $G_{max}$ is an auxiliary variable introduced to represent the maximum contact gap across all pairs of sequentially scheduled contacts. Constraint equation \eqref{eqn:obj_min_max_gap_2} enforces that only one contact can be the next contact after contact $c_i$. Expression \eqref{eqn:obj_min_max_gap_3} forces the auxiliary variable to bound the gap between every sequentially scheduled pairs of contacts. Finally, expressions \eqref{eqn:obj_min_max_gap_4} and \eqref{eqn:obj_min_max_gap_5} enforce that if $y_{ij} = 1$ then both of the associated contacts are also $1$.

\subsection{Constraint Functions}

There are a number of required constraints that must be introduced to ensure that if a collect $c$ is selected, the decision variables of the corresponding provider $p$ and location $l$ are also set. First, we ensure that if a contact is selected, the associated location is also scheduled
\begin{equation}
\begin{aligned}
   	\sum_{(c, C) \in {\mathcal{C}}^L}c \leq |{\mathcal{C}}^L|l, \; \forall \; (l,L) \in {\mathcal{L}}
 \end{aligned}
\label{eqn:constraint_location_selection}
\end{equation}
We enforce a similar constraint so that if a single location for a given provider is selected the provider is selected as well through
\begin{equation}
\begin{aligned}
   	\sum_{(l, L) \in {\mathcal{L}}^P}l \leq |{\mathcal{L}}^P|p, \; \forall \; (p, P) \in {\mathcal{P}}
 \end{aligned}
\label{eqn:constraint_provider_selection}
\end{equation}
Finally, if any contact $c$ for satellite $S$ and station $L$ is scheduled, the associated vehicle indicator variable $v^{S,L}$ must be set to ensure that per-station, per-satellite license costs are counted. This constraint is expressed as
\begin{equation}
\begin{aligned}
   	\sum_{c \in {\mathcal{C}}^{S,L}}c \leq |{\mathcal{C}}^{S,L}|v^{S,L}, \; \forall \; l \in {\mathcal{L}}, s \in {\mathcal{S}}
 \end{aligned}
\label{eqn:constraint_vehicle_indicator}
\end{equation}

It is also possible to introduce a number of different constraint functions that impose systems-engineering requirements on the resulting solution. Not all of these constraints need to be added to the formulation of the problem, but may be included to model a design requirement.

The \textit{minimum constellation data downlink constraint} enforces that the total data volume downlinked across a constellation is above a minimum threshold. This is appropriate for constellations with inter-satellite communications so long as a minimum data downlink volume is achieved that all data can be downlinked across the constellation. The constraint is
\begin{equation}
\begin{aligned}
	& \sum_{(c,C) \in {\mathcal{C}}}C_{dr}C_{duration}c \geq D_{min} \forall \; C \; \text{s.t.} \\
   	& C_{start} \leq t^{window}_{end} \\
   	& C_{end} \geq t^{window}_{start} \\
   	& t^{window}_{end} \leftarrow t^{window}_{start} + T^{period} \\
   	& t^{window}_{start} \in \{t^s_{sim},t^s_{sim}+t_{step},\ldots,t^e_{sim}-T^{period}\}
\end{aligned}
\label{eqn:constraint_min_constellation_data_downlink}
\end{equation}
The constraint enforces that the total data downlink by all contacts across all satellites within a duration of $T^{period}$ exceeds the a minimum threshold $D_{min}$. The constraint is applied in to all contacts in discrete time windows separated by $t_{step}$. This is useful for ensuring that a minimum amount of data is downlinked on a specific cadence such as every orbit or every day.

The \textit{minimum satellite data downlink constraint} enforces that the total data downlinked for each satellite is above a minimum value over any given period $T_{period}$. It is similar to the constellation downlink constraint, however the requirement is applied to each satellite $s \in S$ separately. The constraint is 
\begin{equation}
\begin{aligned}
   	& \sum_{(c,C) \in {\mathcal{C}}^S}C_{dr}C_{duration}c \geq D^S_{min} \forall \; S \; \in {\mathcal{S}}, C \; \text{s.t.} \\
   	& C_{start} \leq t^{window}_{end} \\
   	& C_{end} \geq t^{window}_{start} \\
   	& t^{window}_{end} \leftarrow t^{window}_{start} + T^{period} \\
   	& t^{window}_{start} \in \{t^s_{sim},t^s_{sim}+t_{step},\ldots,t^e_{sim}-T^{period}\}
\end{aligned}
\label{eqn:constraint_min_satellite_data_downlink}
\end{equation}

The \textit{maximum operational cost constraint} ensures that the recurring operational costs from monthly station charges $e^L_{monthly}$ combined with any costs $e^L_{pass}$ and $e^L_{minute}$ associated with taking contacts do not exceed a set monthly threshold $e_{max}$. This can be used for mission budgeting to ensure expected costs stay within a set limit. The constraint is
\begin{equation}
\begin{aligned}
  & \frac{365.25 \times 86400}{12T_{sim}}\sum_{L \in {\mathcal{L}}}\sum_{(c,C) \in {\mathcal{C}}^L}(e^L_{pass}C_{pass} + C_{pass})c \; + \\
  & \hspace{2em} \sum_{(l,L) \in {\mathcal{L}}}e^L_{monthly}l \, \leq E_{max}
\end{aligned}
\label{eqn:constraint_max_operational_cost}
\end{equation}
The constraint is similar to Equation \eqref{eqn:obj_min_cost} though considers only operational costs and is normalized such that the cost is in terms of dollars-per-month.

The \textit{station contact exclusion constraint} ensures that a location can only communicate with one satellite at any time. This enforces the physical constraints on planning that arise from having one antenna at a given site. This could be modified to limit the number of antennas up to a set number, though we only consider a single antenna per location in this work. This constraint should be added to most problems. The constraint is
\begin{equation}
\begin{aligned}
   	& c_i + c_j \leq 1 \; \\
   	& \hspace{2em} \forall \; L \in {\mathcal{L}}, i,j \in \{0,\ldots,|{\mathcal{C}}^L|\}, j > i \; \text{s.t.} \\
   	& C_{start,i} \leq C_{end,j} \\
   	& C_{start,j} \leq C_{end,j} \\
\end{aligned}
\label{eqn:constraint_station_contact_exclusion}
\end{equation}

The \textit{satellite contact exclusion constraint} ensures that a satellite can only communicate with a single location at any time. This constraint should be added to most problems unless the satellites can support simultaneous contacts with multiple locations simultaneously. The constraint can be expressed as
\begin{equation}
\begin{aligned}
	& c_i + c_j \leq 1 \; \\
   	& \hspace{2em} \forall \; S \in {\mathcal{S}}, i,j \in \{0,\ldots,|{\mathcal{C}}^S|\}, j > i \; \text{s.t.} \\
   	& C_{start,i} \leq C_{end,j} \\
   	& C_{start,j} \leq C_{end,j} \\
\end{aligned}
\label{eqn:constraint_contact_exclusion}
\end{equation}

The \textit{maximum contact gap constraint} ensures that the maximum time gap between any two contacts is less than a specific duration. It is similar to the \textit{minimum maximum contact gap objective} only instead of trying to minimize the gap, ensures that a requirement on the gap is met. This constraint is expressed as 
\begin{equation}
\begin{aligned}
   	& (C_{start,j} - C_{end,i})y_{ij} \leq G_{max}, \; \forall \; S \in {\mathcal{S}} \\
   	& \hspace{2em} (c,C) \in {\mathcal{C}}^S, i,j \in \{0,\ldots,|{\mathcal{C}}^S|\} \\
   	& \sum_{j>i}y_{ij} = c, \\    
    & y_{ij} \leq c_i, \\
    & y_{ij} \leq c_j, \\
    & y_{ij} \in \{0,1\} \\
\end{aligned}
\label{eqn:constraint_contact_gap}
\end{equation}

%\begin{align}
%    &\underset{\mathbf{x}}{\text{minimize}} && G_{max} \label{eqn:obj_min_max_gap_1} \\ 
%    &\text{subject to} && \sum_{j>i}y_{ij} = c, \forall \; (c,C) \in {\mathcal{C}}^S, S \in {\mathcal{S}}, \label{eqn:obj_min_max_gap_2} \\ % \forall i \in |X^{p,l}|
%    & && \hspace{2em} i,j \in \{0,\ldots,|{\mathcal{C}}^S|\}\; \text{s.t.} \\
%    & && \; C_{start,j} > C_{start,j} \\
%    & && (C_{start,j} - C_{end,i})y_{ij} \leq G_{max},  \label{eqn:obj_min_max_gap_3}\\
%    & && y_{ij} \leq c_i, \label{eqn:obj_min_max_gap_4}\\
%    & && y_{ij} \leq c_j, \label{eqn:obj_min_max_gap_5}\\
%    & && y_{ij} \in \{0,1\}, \label{eqn:obj_min_max_gap_6}\\
%    & && c \in \{0,1\},  \label{eqn:obj_min_max_gap_7}
%\end{align}

The \textit{maximum provider constraint} ensures that the number of ground station providers selected is at most $P_{max}$
\begin{equation}
\begin{aligned}
   	\sum_{(p,P) \in {\mathcal{P}}} p \leq P_{max}
\end{aligned}
\label{eqn:constraint_max_proivder}
\end{equation}

The \textit{minimum contact duration constraint} ensures that only contacts of with a duration greater than $t_{min}$ are considered for planning. This constraint eliminates short-duration contacts that might not be operationally useful
\begin{equation}
\begin{aligned}
   	& x = 0 \; \text{if} \; C_{end} - C_{start} < t_{min}, \forall \; (c,C) \in {\mathcal{C}}
\end{aligned}
\label{eqn:constraint_min_duration}
\end{equation}

The \textit{minimum contacts-per-period} constraint ensures that each satellite takes at least $N_{min}$ contacts over any given period of duration $T^{period}$. The constraint is enforced over discrete periods with a an incremental offset of $t_{step}$ over the simulation window
\begin{equation}
\begin{aligned}
   	& \sum_{(c,C) \in {\mathcal{C}}}c \geq N_{min}, \; \forall \; C \; \text{s.t.}\\
   	& C_{start} \leq t^{window}_{end} \\
   	& C_{end} \geq t^{window}_{start} \\
   	& t^{window}_{end} \leftarrow t^{window}_{start} + T^{period} \\
   	& t^{window}_{start} \in \{t^s_{sim},t^s_{sim}+t_{step},\ldots,t^e_{sim}-T^{period}\}
\end{aligned}
\label{eqn:constraint_minimum_contacts_per_period}
\end{equation}

The \textit{required provider constraint} ensures that a specific provider $P$ is selected
\begin{equation}
\begin{aligned}
   	p = 1
\end{aligned}
\label{eqn:constraint_required_provider}
\end{equation}

The \textit{required location constraint} ensures that location $L$ is selected
\begin{equation}
\begin{aligned}
   	l = 1
\end{aligned}
\label{eqn:constraint_required^location}
\end{equation}

The \textit{station number constraint} ensures that at least $M_{min}$ stations are selected and no more than $M_{max}$ are selected
\begin{equation}
\begin{aligned}
   	& \sum_{l \in {\mathcal{L}}} l \geq M_{min} \\
   	& \sum_{l \in {\mathcal{L}}} l \leq M_{max} \\
\end{aligned}
\label{eqn:constraint_required^location}
\end{equation}

If a constraint on the \textit{station frequencies} is required to ensure compatibility with satellite communications design, such as supporting $S$-band uplink and $X$-band downlink, the constraint can be implemented by pre-filtering the set of all locations $\mathcal{L}$ and only considering those that support the required frequencies. Implementing this constraint by pre-filtering is better than introducing a decision variable, this reduces the number of decision variables and improves performance. All non-compatible ground stations would not be scheduled regardless, so there is no loss of generality with this approach.

%% file: sections/experiments.tex
\section{Experiments}
\label{sec:experiments}

% Know hard-coded ref is bad, but this document class doesn't like \Cref

To evaluate the performance of the IP optimization solution, we simulate optimization problems by randomizing the cost and data downlink parameters across all providers and stations in Appendix A. The range of potential values used to generate the random scenarios are listed in \Cref{tab:constant_settings}. Spacecraft are randomly sampled from the set of objects with altitudes between \SI{300}{\kilo\meter} to \SI{1000}{\kilo\meter} from the Celestrak active satellite database \cite{celestrak}.
\begin{table}[ht]
\centering
\begin{tabular}{lr}
\toprule
Parameter & Range \\
\midrule
$e^P_{integ}$ & [50000, 200000] \\
$e^L_{setup}$ & [10000, 100000] \\
$e^L_{monthly}$ & [200, 5000] \\
$e^L_{license}$ & [1000, 5000] \\
$e^L_{pass}$ & [25, 175] \\
$e^L_{minute}$ & [5, 35] \\
$S_{dr}$ & [$9 \times 10^8$, $1.8 \times 10^9$] \\
$L_{dr}$ & [$1.2 \times 10^9$, $1.8 \times 10^9$] \\
\bottomrule
\end{tabular}
\caption{Distribution of constant values for scenarios} 
\label{tab:constant_settings}
\end{table}

%We consider the three variations of the ground station optimization problem outlined in \Cref{sec:problem_variations}.

 The spacecraft dynamics are propagated using the SGP4 propagator \cite{vallado2006revisiting}. The simulation window $T_{sim}$ is set to be 7 days long and the optimization window $T_{opt}$ is set to 365 days for all cases. The simulation window was selected by propagating the spacecraft dynamics out for various simulation windows and analyzing at what point the number of contacts per day and mean gap and contact duration level out. These results are shown in Appendix \ref{sec:appendix_prediction_assumption}. A $10^\circ$ minimum elevation mask is applied at all locations when calculating contact opportunities. All simulations were run on a workstation with a 64-core 2.0 GHz Intel AMD EPYC 7713 processor.
 
 We consider two primary problem variations\textemdash a total mission cost minimization problem and a data downlink maximization problem. These problem variations are constructed to represent practical system design trade studies that might be undertaken as part of designing a space mission ground segment. The problem variations are constructed by composing an objective function with a subset of the possible constraint functions in \Cref{sec:problem_formulation}.
\begin{table}[ht]
\centering
\begin{tabular}{lr}
\toprule
Parameter & Range \\
\midrule
$t^s_{sim}$ & 2024-09-11 00:00:00 UTC \\
$t^e_{sim}$ & 2024-09-18 00:00:00 UTC \\
$t^s_{opt}$ & 2024-09-11 00:00:00 UTC \\
$t^e_{opt}$ & 2025-09-11 00:00:00 UTC \\
$D^s_{min}$ & $10^{11}$ \\
$T^{period}$ & 86400 \\
$t_{step}$ & 3600 \\
$t_{min}$ & 180 \\
$E_{max}$ & $10^6$\\
\bottomrule
\end{tabular}
\caption{Design parameters used for simulation} 
\label{tab:design_parameters}
\end{table}

To evaluate the IP formulation we performed 100 trials for constellations of 1, 2, 5, and 10 satellites for each type of optimization problem. In each trial, we generate a new problem scenario by randomly setting the simulation parameters by uniformly sampling the values in \Cref{tab:constant_settings} for each provider. We then solve the IP formulation with the Gurobi v11.0.3 optimizer. To provide a solution baseline, we calculate the optimal solution for the restricted problem of location selected for a one or two providers to mimic the behavior of a mission operator integrating with one or two GSaaS providers to support a mission. We then compare the performance of the IP-optimized solution to the these fixed solutions.
 
\subsection{Mission Cost Minimization Problem}

The mission cost minimization problem variation seeks to minimize the total cost of the ground station network over the mission duration, while also enforcing a minimum per-satellite data downlink requirement is met. The problem formulation includes station contact exclusion and satellite contact exclusion constraints. A minimum contact duration constraint is added to ensure short-duration contacts are excluded. A minimum per-satellite data downlink constraint is required to ensure a non-degenerate solution, as utilizing no locations, and no providers will otherwise result in an ``optimal'' zero-cost solution. The full set of equations defining the problem formulation can be found in Appendix \ref{sec:appendix_min_cost_opt}. The values for the constraint design parameters used in the simulations are listed in \Cref{tab:design_parameters}.

\begin{figure*}[ht!]
  \centering
  % Subfigure 1
  \begin{subfigure}{0.3\textwidth}
    \centering
    \includegraphics[width=\textwidth]{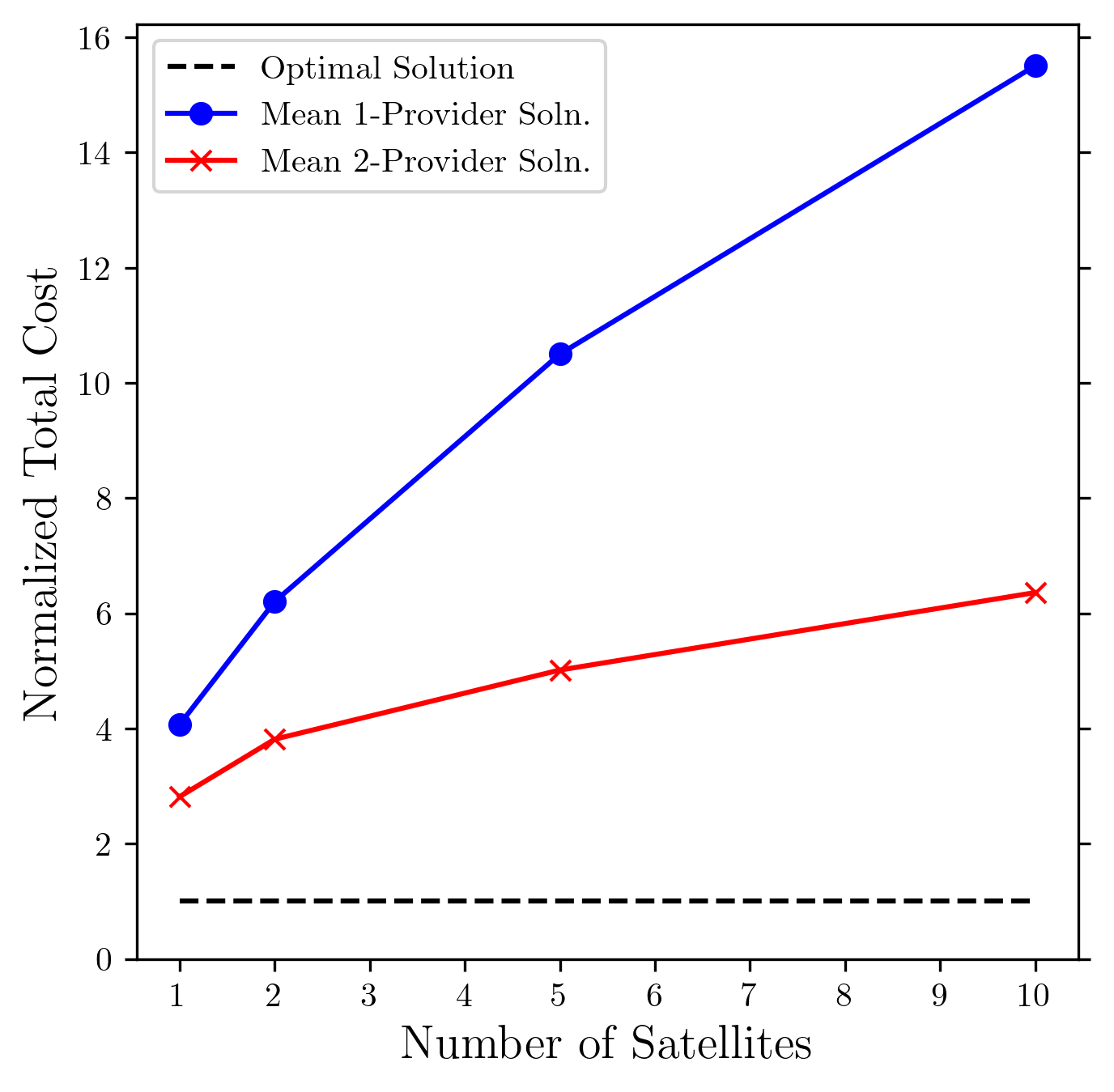}
    \label{fig:sub1}
  \end{subfigure}
  \hfill
  % Subfigure 2
  \begin{subfigure}{0.3\textwidth}
    \centering
    \includegraphics[width=\textwidth]{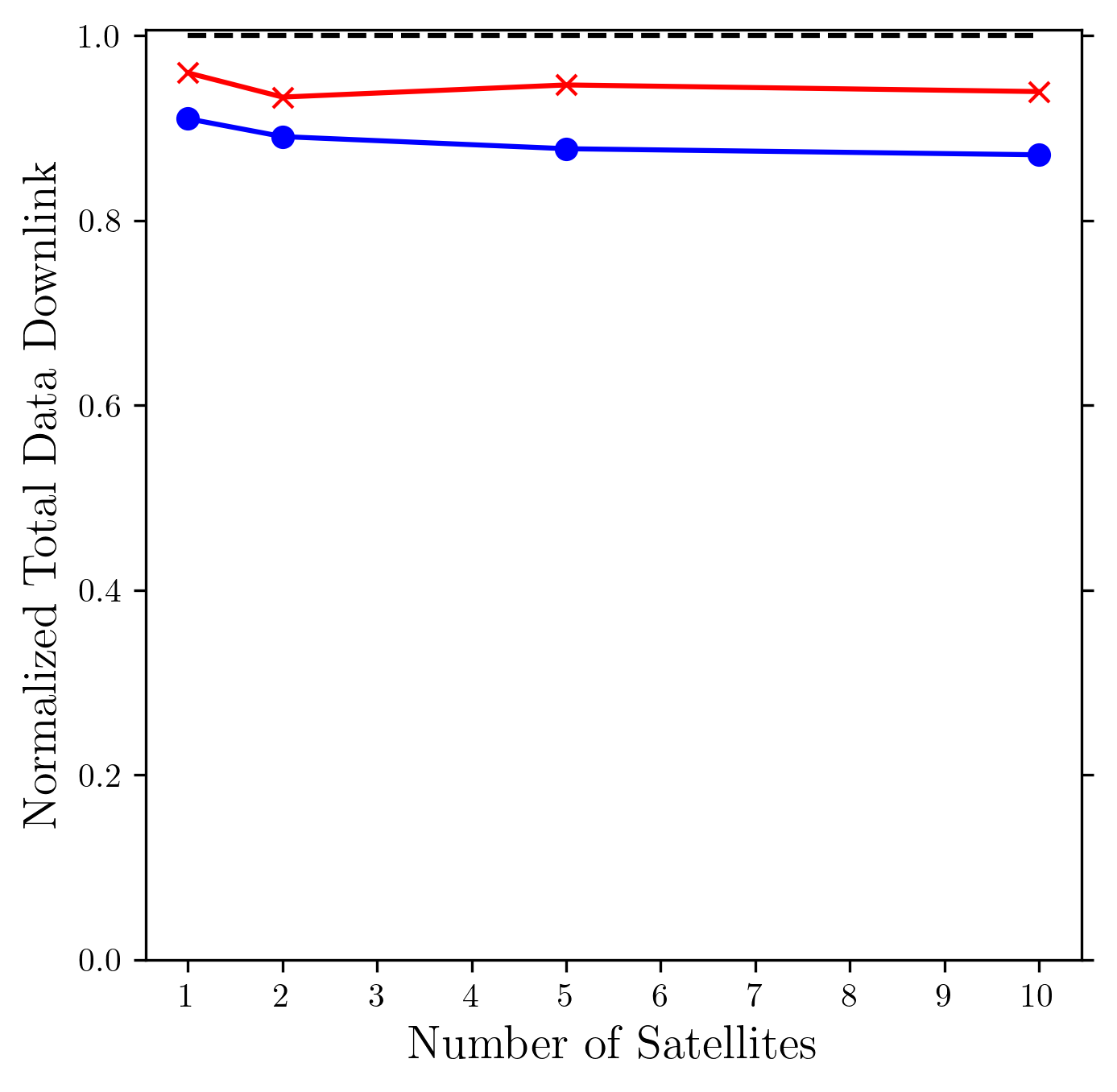} % Replace with your image
%    \caption{Second Subfigure}
    \label{fig:sub2}
  \end{subfigure}
  \hfill
  % Subfigure 3
  \begin{subfigure}{0.3\textwidth}
    \centering
    \includegraphics[width=\textwidth]{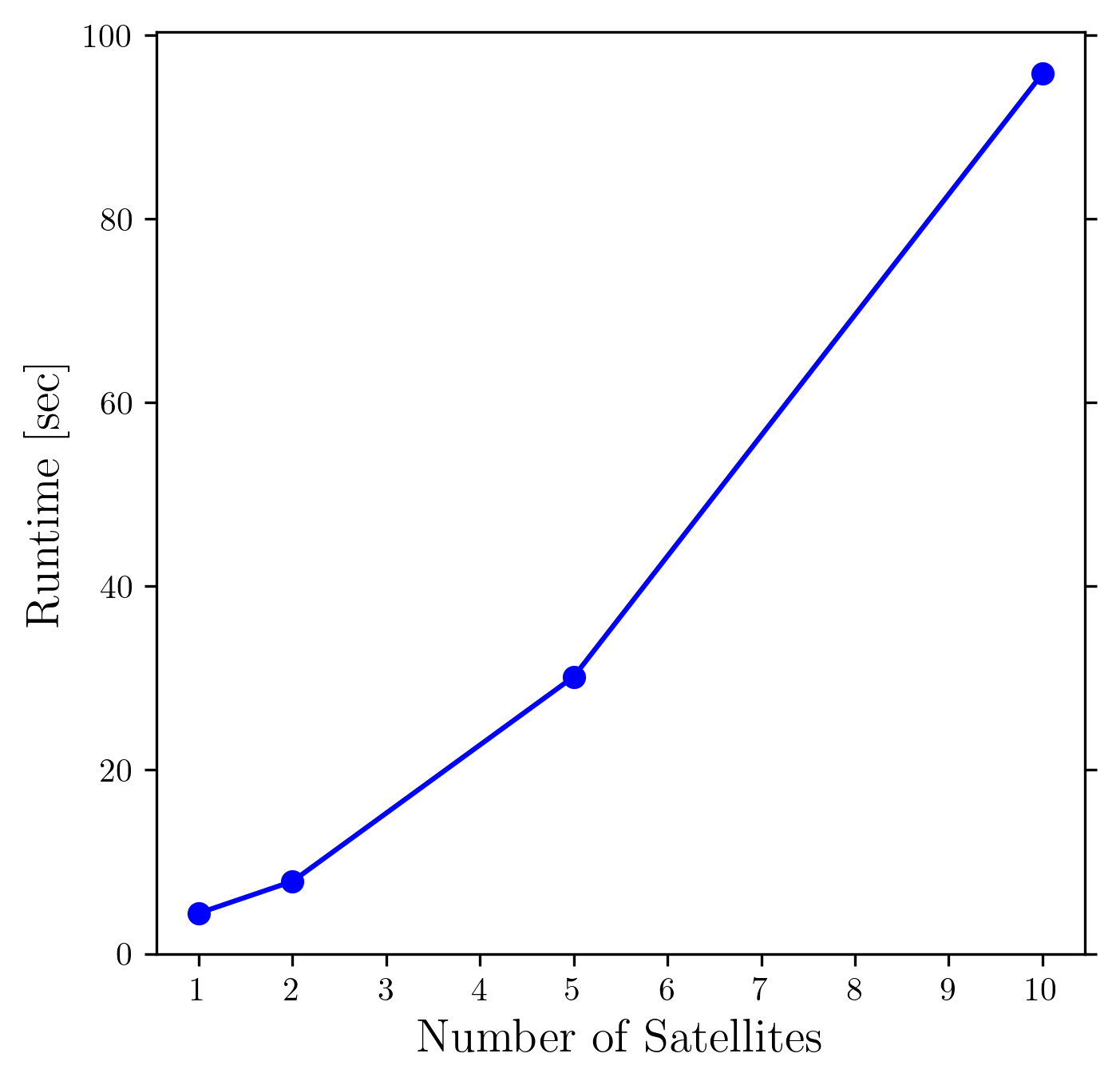} % Replace with your image
%    \caption{Third Subfigure}
    \label{fig:sub3}
  \end{subfigure}
  \caption{Performance comparisons of IP-optimized ground station network with one- and two-provider solutions. Normalized total cost (left), normalized total data downlink (middle) and average optimizer runtime (right). The normalized value is the value of a solution divided by the value of the optimal solution to that scenario. Closer to 1 is better for all normalized values.}
  \label{fig:min_cost_perf}
\end{figure*}

\Cref{fig:min_cost_perf} shows how the optimal solution of the cost minimization problem compares to the average one-provider and two-provider solutions. As expected, the IP optimization problem that is free to optimize over all potential ground station providers out performs the best possible networks when only a one or two ground station providers are considered. One surprising result is that, despite minimizing the total cost, the IP optimzied solution also outperforms the one- and two-provider solutions in terms of total data downlinked at as well, which would not normally be expected since the objective does not include any explicit regularization term that would encourage data downlink maximization. We can see that the IP formulation is relatively fast with the average solution time being under 100 seconds for constellations of up to 10 satellites. One interesting trend that is observed is that the total normalized cost appears to plateau for the two-provider solution when the constellation size reaches 10 satellites. This can be explained by saturation of the ground station network and an inability to support more contacts for new satellites. That is, the constellation size has become large enough that all antennas are in nearly constant use, saturating the network, so that adding more satellites does not lead to an increase in number of contacts taken and therefore added cost.

\subsection{Mission Data Maximization Problem}

The mission data maximization problem seeks to maximize the total amount of data downlinked over the mission, while also ensuring the solution adheres to a maximum operational cost constraint. The problem formulation includes station contact exclusion and satellite contact exclusion constraints. A minimum contact duration constraint is also included. A maximum operational cost constraint is required to ensure that the optimizer does not simply select all providers and locations to provide the most potential downlink opportunities resulting in a degenerate solution. The design constants of this problem are $E_{max}$, the maximum allowed monthly operational cost and $t_{min}$, the minimum acceptable contact duration. The full set of equations defining the problem formulation can be found in Appendix \ref{sec:appendix_max_data_opt}. The design parameters used in the simulations are listed in \Cref{tab:design_parameters}.

\begin{figure*}[ht!]
  \centering
  % Subfigure 1
  \begin{subfigure}{0.3\textwidth}
    \centering
	\includegraphics[width=\textwidth]{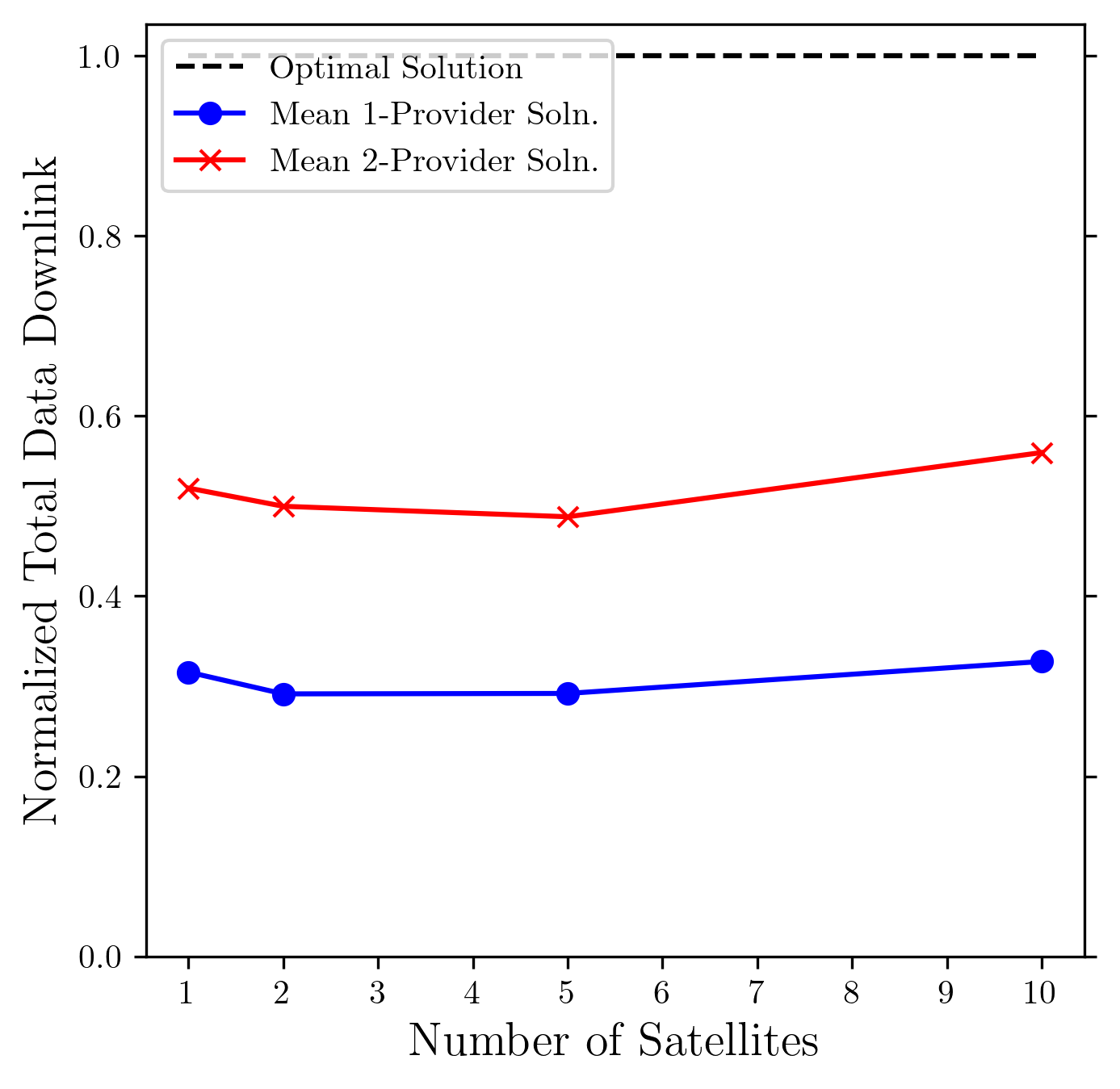}
%    \caption{First Subfigure}
  \label{fig:sub1}
  \end{subfigure}
  \hfill
  % Subfigure 2
  \begin{subfigure}{0.3\textwidth}
    \centering
    \includegraphics[width=\textwidth]{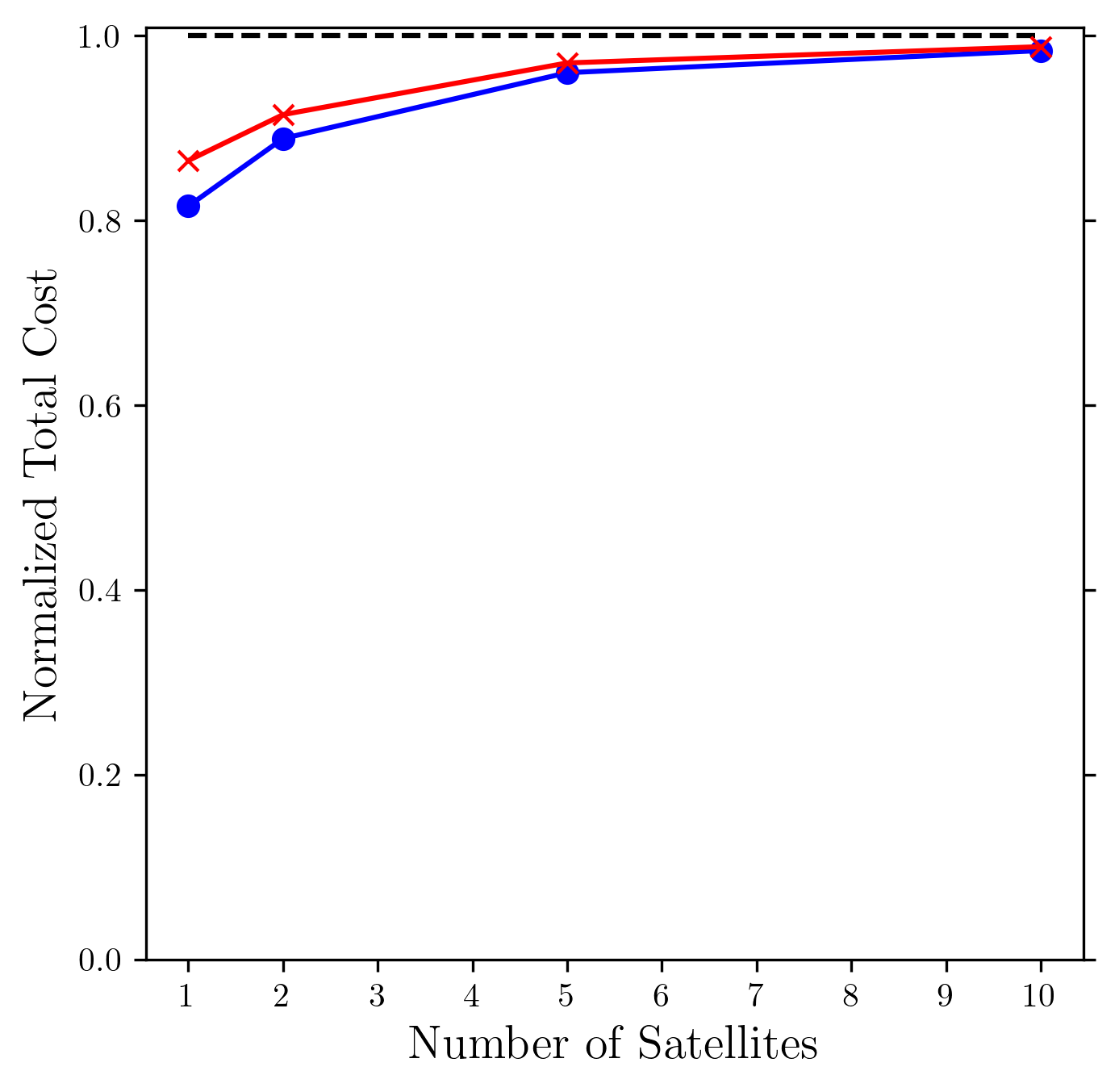} % Replace with your image
%    \caption{Second Subfigure}
    \label{fig:sub2}
  \end{subfigure}
  \hfill
  % Subfigure 3
  \begin{subfigure}{0.3\textwidth}
    \centering
    \includegraphics[width=\textwidth]{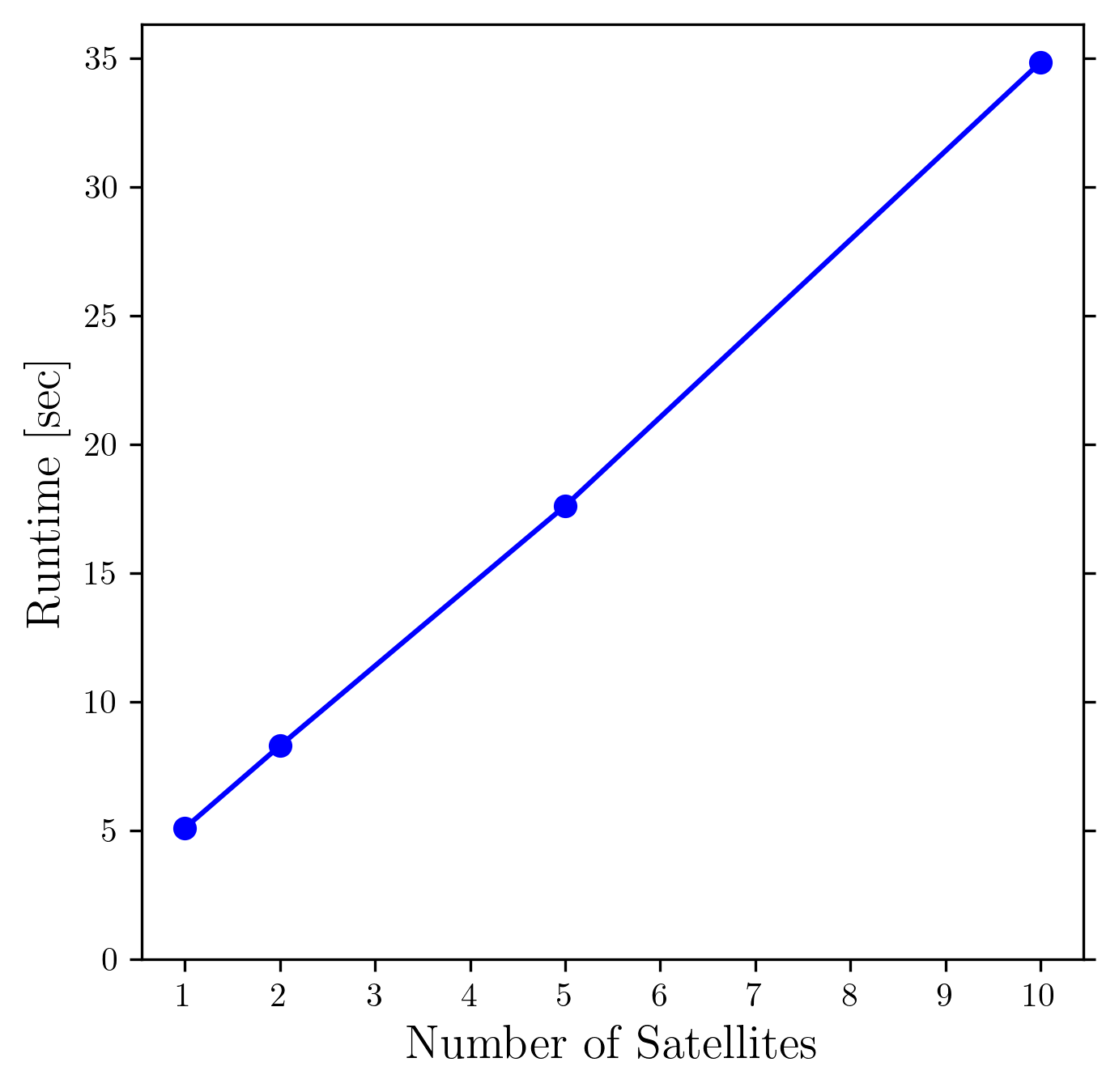} % Replace with your image
%    \caption{Runtime}
    \label{fig:sub3}
  \end{subfigure}
  \caption{Performance comparisons of IP-optimized ground station network with one- and two-provider solutions. Normalized total data downlink (left), normalized total cost (middle) and average optimizer runtime (right). The normalized value is the value of a solution divided by the value of the optimal solution to that scenario. Closer to 1 is better for the objective value and worse for cost.}
  \label{fig:max_data_perf}
\end{figure*}

\Cref{fig:max_data_perf} shows that optimizing a ground station network for maximal data downlink can significantly increase the total amount of data throughput compared to one- and two-provider ground station networks without significantly increasing cost. For a 1 satellite mission the data-optimized ground station network is able to downlink over 300\% more data over the entire life of the mission at only 25\% increase cost compared to a one-provider solution, and able to downlink over 200\% more data compared to a two-provider solution. For a 10 satellite constellation the optimized ground station network can still downlink nearly 300\% and 200\% more data than the one- and two-provider networks at only 5\% increased cost. The explanation for this significant performance improvement is that the IP formulation is able to use the lowest-cost station in regions where there is overlapping station coverage (e.g. Europe), as well as add single stations in non-saturated regions (e.g. the Pacific Ocean) to increase total capacity. The runtime of the solution increases nearly linearly with constellation size with the optimal network for a 10-satellite constellation being found within 35 seconds on average.

%\begin{figure*}[ht!]
%  \centering
%  % Subfigure 1
%  \begin{subfigure}{0.245\textwidth}
%    \centering
%    \includegraphics[width=\textwidth]{figures/sims/max_data_opt_objective_trials_sats_1.png} % Replace with your image
%%    \caption{First Subfigure}
%  \label{fig:sub1}
%  \end{subfigure}
%  \hfill
%  % Subfigure 2
%  \begin{subfigure}{0.245\textwidth}
%    \centering
%    \includegraphics[width=\textwidth]{figures/sims/max_data_opt_objective_trials_sats_2.png} % Replace with your image
%%    \caption{Second Subfigure}
%    \label{fig:sub2}
%  \end{subfigure}
%  \hfill
%  % Subfigure 3
%  \begin{subfigure}{0.245\textwidth}
%    \centering
%    \includegraphics[width=\textwidth]{figures/sims/max_data_opt_objective_trials_sats_5.png} % Replace with your image
%%    \caption{Third Subfigure}
%    \label{fig:sub3}
%  \end{subfigure}
%  % Subfigure 4
%  \begin{subfigure}{0.245\textwidth}
%    \centering
%    \includegraphics[width=\textwidth]{figures/sims/max_data_opt_objective_trials_sats_10.png} % Replace with your image
%%    \caption{Third Subfigure}
%    \label{fig:sub4}
%  \end{subfigure}
%  \caption{Optimal normalized data downlink of average 1-provider and 2-provider solutions across 100 randomly generated scenarios for constellations of 1 (outer left), 2 (middle left), 5 (middle right), and 10 (outer right) satellites.}
%  \label{fig:max_data_trials}
%\end{figure*}

%% file: sections/conclusions.tex
\section{Conclusions}
\label{sec:conclusions}

This paper introduced and solved the problem of ground station provider and location selection for space mission designers. The problem was posed as an integer programming problem. Three different objective functions and fourteen different potential constraint functions were presented. These objective functions and constraint functions can be composed to conduct different types of design studies of satellite ground station networks. Two primary classes of optimization problems were analyzed\textemdash mission cost minimization and data volume maximization. For mission cost minimization, the IP solution was shown to reduce total mission cost while simultaneously improving total data downlink. For data volume maximization problems, an IP-optimized ground station network was seen to increase the total data downlink by up to 300\% compared to optimal single-provider networks and 200\% compared to the optimal two provider networks. In future work, it would be desirable to consider a multi-objective optimization of ground networks so that instead of optimizing for a single design parameter, the network is optimized simultaneously for multiple considerations. Additionally, it would be desirable to investigate the unconstrained network design problem, where instead of selecting from a set of fixed locations, a network is optimized over any viable terrestrial location. One area of potential improvement is that the max gap objective and constraint introduce a large number of auxiliary variables and equations which has a significant negative impact on performance, therefore finding an alternative formulation or solution approach is desirable. 

%% file: sections/appendix_gs_list.tex
\section{Commercial GSaaS Station List}
\label{sec:appendix_station_list}

A list of current major commercial GSaaS locations is provided in \Cref{tab:ground_stations}. Please note that due to security concerns these are not the exact station coordinate. All coordinates are provided to two decimal places of accuracy at the request of the participating providers. For non-participating providers, the station locations was obtained by correlating publicly available data with the provider's stated ground network. This accuracy is sufficient for most analysis purposes.

\begin{table*}[ht]
\centering

\caption{List of current ground stations from major commercial GSaaS Providers}
\begin{tabular}{llrrcccccc}
%\specialrule{.1em}{.05em}{.2em}
\toprule
\multicolumn{4}{l}{\phantom{-}} & \multicolumn{5}{c}{Supported Frequency Bands} & \phantom{-} \\
\midrule
Location & Country & Longitude [deg] & Latitude [deg] & $UHF$ & $L$ & $S$ & $X$ & $Ka$ & Status \\
\midrule
\multicolumn{10}{l}{\textbf{Atlas Space Operations$^{\star}$}\cite{atlasspace}} \\
Awarua & New Zealand & 168.38 & -46.52 & & & $\times$ & $\times$ & & Operational \\
Chitose & Japan & 141.62 & 42.77 & & & $\times$ & $\times$ & & Operational \\
Dubai & United Arab Emirates & 55.35 & 24.94 & & & $\times$ & $\times$ & & Operational \\
Dundee & Scotland & -3.18 & 56.40 & & & $\times$ & $\times$ & & Operational \\
Harmon & Guam & 144.82 & 13.51 & & & $\times$ & $\times$ & & Operational \\
Mojave & USA & -118.15 & 35.05 & & & $\times$ & $\times$ & & Operational \\
Mwulire & Rwanda & 30.39 & -1.96 & & & $\times$ & $\times$ & & Operational \\
Paumalu & USA & -158.03 & 21.67 & & & $\times$ & $\times$ & & Operational \\
Sunyani & Ghana & 55.35 & 24.94 & & & $\times$ & & & Operational \\
Tahiti & French Polynesia & -149.60 & -17.63 & & & $\times$ & $\times$ & & Operational \\
Utqiagvik & USA & -156.81 & 71.27 & & & $\times$ & $\times$ & & Operational \\

\multicolumn{10}{l}{\textbf{Amazon Web Services Ground Station$^{\star}$}\cite{awsgs,capellantia}} \\
Alaska & USA & -149.90 & 71.29 & & & $\times$ & $\times$ & & Operational \\
Cape Town & South Africa & 18.43 & -34.01 & & & $\times$ & $\times$ & & Operational \\
Dubbo & Australia & 148.63 & -32.18 & & & $\times$ & $\times$ & & Operational \\
Dublin & Ireland & -6.13 & 53.24 & & & $\times$ & $\times$ & & Operational \\
Hawaii & USA & -157.84 & 21.32 & & & $\times$ & $\times$ & & Operational \\
Manama & Bahrain & 50.30 & 26.03 & & & $\times$ & $\times$ & & Operational \\
Ohio & USA & -83.12 & 40.06 & & & $\times$ & $\times$ & & Operational \\
Oregon & USA & -119.37 & 45.51 & & & $\times$ & $\times$ & & Operational \\
Punta Arenas & Argentina & -70.87 & -52.94 & & & $\times$ & $\times$ & & Operational \\
Seoul & South Korea & 127.02 & 36.74 & & & $\times$ & $\times$ & & Operational \\
Singapore & Singapore & 103.70 & 1.31 & & & $\times$ & $\times$ & & Operational \\
Stockholm & Sweden & 16.35 & 59.39 & & & $\times$ & $\times$ & & Operational \\

\multicolumn{10}{l}{\textbf{Azure Orbital Ground Station$^{\star}$}\cite{azureorbital}} \\

Galve & Sweden & 17.34 & 60.63 & & & $\times$ & $\times$ & & Decomissioned \\
Johannesburg & South Africa & 28.20 & -26.14 & & & $\times$ & $\times$ & & Decomissioned \\
Longovilo & Chile & -71.40 & -33.96 & & & $\times$ & $\times$ & & Decomissioned \\
Quincy & USA & -119.89 & 47.24 & & & $\times$ & $\times$ & & Decomissioned \\
Singapore & Singapore & 103.85 & 1.28 & & & $\times$ & $\times$ & & Decomissioned \\

\multicolumn{10}{l}{\textbf{Kongsberg Satellite Services$^{\dagger}$}} \\

Awarua & New Zealand & 168.38 & -46.53 & & & $\times$ & $\times$ & $\times$ & Operational \\
Azores & Azores & -25.13 & 36.99 & & & $\times$ & $\times$ & & Operational \\
Bangalore & India & 77.37 & 12.9 & & & $\times$ & $\times$ & & Operational \\
Bel Ombre & Mauritius & 57.45 & -20.5 & & & $\times$ & $\times$ & & Operational \\
Cordoba & Argentina & -66.1 & -33.2 & & & $\times$ & & & Operational \\
Dubai & United Arab Emirates & 55.47 & 25.23 & & & $\times$ & $\times$ & & Operational \\
Fairbanks & USA & -147.72 & 64.82 & & & $\times$ & $\times$ & $\times$ & Operational \\
Hartebeesthoek & South Africa & 27.71 & -25.89 & & & $\times$ & $\times$ & & Operational \\
Hawaii & USA & -156.45 & 20.82 & & & $\times$ & $\times$ & & Operational \\
Hokkaido & Japan & 143.45 & 42.6 & & & $\times$ & $\times$ & & Operational \\
Inuvik & Canada & -133.61 & 68.33 & & & $\times$ & $\times$ & & Operational \\
Jan Mayen & Norway & -8.72 & 70.92 & & & $\times$ & $\times$ & & Operational \\
Jeju & South Korea & 126.32 & 33.39 & & & $\times$ & $\times$ & & Operational \\
Kourou & French Guiana & -52.8 & 5.25 & & & $\times$ & $\times$ & & Operational \\
Long Beach & USA & -118.15 & 33.82 & & & $\times$ & $\times$ & & Operational \\
Maspalomas & Spain & -15.64 & 27.77 & & & $\times$ & $\times$ & $\times$ & Operational \\
Miami & USA & -80.38 & 25.61 & & & $\times$ & $\times$ & & Operational \\
Mingenew & Australia & 115.34 & -29.01 & & & $\times$ & $\times$ & & Operational \\
Nemea & Greece & 22.62 & 37.85 & & & $\times$ & $\times$ & & Operational \\
Nuuk & Greenland & -51.7 & 64.19 & & & $\times$ & $\times$ & & Operational \\
Okinawa & Japan & 127.81 & 26.36 & & & $\times$ & $\times$ & & Operational \\
Peterborough & Australia & 138.86 & -32.96 & & & $\times$ & $\times$ & & Operational \\
Prudhoe Bay & USA & -148.47 & 70.2 & & & $\times$ & $\times$ & & Operational \\
Puertollano & Spain & -4.16 & 38.67 & & & $\times$ & $\times$ & & Operational \\
Punta Arenas & Chile & -70.87 & -52.94 & & & $\times$ & $\times$ & $\times$ & Operational \\
Singapore & Singapore & 103.7 & 1.32 & & & $\times$ & $\times$ & & Operational \\
Svalbard & Norway & 15.41 & 78.23 & & & $\times$ & $\times$ & $\times$ & Operational \\
Thermopylae & Greece & 22.69 & 38.82 & & & $\times$ & $\times$ & & Operational \\
Thomaston & USA & -84.26 & 32.95 & & & $\times$ & $\times$ & & Operational \\
Tokyo & Japan & 139.35 & 36.0 & & & $\times$ & $\times$ & & Operational \\
Tolhuin & Argentina & -67.12 & -54.51 & & & $\times$ & & & Operational \\

\bottomrule
\end{tabular}
\label{tab:ground_stations}
\end{table*}

\begin{table*}[ht]
\centering
%\caption{List of current ground stations from major commercial GSaaS Providers}
\begin{tabular}{llrrcccccc}
\toprule
\multicolumn{4}{l}{\phantom{-}} & \multicolumn{5}{c}{Supported Frequency Bands} & \phantom{-} \\
\midrule
Location & Country & Longitude [deg] & Latitude [deg] & $UHF$ & $L$ & $S$ & $X$ & $Ka$ & Status \\
\specialrule{.05em}{.05em}{.05em}

\multicolumn{10}{l}{\textbf{Kongsberg Satellite Services (Cont.)}} \\
Troll & Antarctica & 2.53 & -72.0 & & & $\times$ & $\times$ & $\times$ & Operational \\
Tromso & Norway & 18.95 & 69.66 & & & $\times$ & $\times$ & & Operational \\
Tuskuba & Japan & 140.08 & 36.13 & & & $\times$ & $\times$ & & Operational \\
Vardo & Norway & 31.03 & 70.34 & & & $\times$ & & & Operational \\
Weilheim & Germany & 11.08 & 47.88 & & & $\times$ & $\times$ & & Operational \\
Accra & Ghana & -0.23 & 5.67 & & & $\times$ & $\times$ & & Potential \\
Brasilia & Brazil & -47.89 & -15.80 & & & $\times$ & $\times$ & & Potential \\
Longovilo & Chile & -71.40 & -33.96 & & & $\times$ & $\times$ & & Potential \\
Maine & USA & -67.89 & 46.95 & & & $\times$ & $\times$ & & Potential \\
New Mexico & USA & -106.91 & 32.27 & & & $\times$ & $\times$ & & Potential \\
San Jose & Costa Rica & -84.06 & 9.95 & & & $\times$ & $\times$ & & Potential \\

\multicolumn{10}{l}{\textbf{Leaf Space$^{\dagger}$}\cite{leafline}} \\
Absheron & Azerbaijan & 40.49 & 40.47 & & & $\times$ & $\times$ & & Operational \\
Awarua & New Zealand & 168.38 & -46.53 & & & $\times$ & $\times$ & $\times$ & Operational \\
Blonduos & Iceland & -20.24 & 65.65  & & & $\times$ & $\times$ & $\times$ & Operational \\
Jeju & South Korea & 126.32 & 33.39 & & & $\times$ & $\times$ & & Operational \\
Kandy & Sri Lanka & 80.72 & 7.27 & & & $\times$ & $\times$ & & Operational \\
Kaspichan & Bulgaria & 27.16 & 43.31 & $\times$ & & $\times$ & & & Operational \\
La Paz & Mexico & -110.39 & 24.10 & & & $\times$ & $\times$ & & Operational \\
Lomazzo & Italy & 9.04 & 45.70 & $\times$ & & & & & Operational \\
Mon Loisir & Mauritius & 57.87 & -20.14 & $\times$ & & $\times$ & $\times$ & & Operational \\
Nangetty & Australia & 115.34 & -29.01 & & & $\times$ & $\times$ & & Operational \\
Peterborough & Australia & 138.85 & -32.96 & & & $\times$ & $\times$ & & Operational \\
Plana & Bulgaria & 23.45 & 42.48 & & & $\times$ & $\times$ & & Operational \\
Pretoria & South Africa & 28.45 & 25.86 & & & $\times$ & $\times$ & & Operational \\
Puetrollano & Spain & -4.16 & 38.67  & $\times$ & & $\times$ & & & Operational \\
Punta Arenas & Chile & -70.85 & -53.04 & & & $\times$ & $\times$ & $\times$ & Operational \\
Santa Maria & Azores & -25.14 & 37.00 & $\times$ & & $\times$ & $\times$ & & Operational \\
Unst & United Kingdom & -0.86 & 60.75 & $\times$ & & $\times$ & $\times$ & & Operational \\
Vimercate & Italy & 9.36 & 45.59 & $\times$ & & $\times$ & & & Operational \\
Longwood & St Helena & -5.66 & -15.94 & & & $\times$ & $\times$ & & Planned \\
Maui & USA & -156.44 & 20.75 & & & $\times$ & $\times$ & & Planned \\
Montsec & Spain & 0.73 & 42.05 & & & $\times$ & $\times$ & $\times$ & Planned \\
Nova Scotia & Canada & -61.04 & 45.32 & & & $\times$ & $\times$ & & Planned \\
Santiago & Chile & -70.77 & 33.36 & & & $\times$ & $\times$ & & Planned \\
Talkeetna & USA & -150.03 & 62.33 & & & $\times$ & $\times$ & & Planned \\
Utqiagvik & USA & -156.8 & 71.28 & & & $\times$ & $\times$ & $\times$ & Planned \\

\multicolumn{10}{l}{\textbf{Viasat Real-Time Earth$^{\dagger}$}\cite{viasatrte}} \\

Accra & Ghana & -0.30 & 5.60 & & $\times$ & $\times$ & $\times$ & $\times$ & Operational \\
Alice Springs & Australia & 133.88 & -23.76 & & $\times$ & $\times$ & $\times$ & $\times$ & Operational \\
Cordoba & Argentina & -64.46 & -31.52 & & & $\times$ & $\times$ & & Operational \\
Fairbanks & USA & -147.54 & 64.79 & & & $\times$ & $\times$ & $\times$ & Operational \\
Guildford & United Kingdom & -0.62 & 51.24 & & & $\times$ & $\times$ & & Operational \\
Hokkaido & Japan & 143.45 & 42.59 & & & $\times$ & $\times$ & $\times$ & Operational \\
Krugersdorp & South Africa & 27.70 & 25.88 & & & $\times$ & $\times$ & $\times$ & Operational \\
Pendergrass & USA & -83.67 & 34.18 & & & $\times$ & $\times$ & & Operational \\
Pieta & Sweden & -0.30 & 5.60 & & & & $\times$ & & Operational \\

\bottomrule
\end{tabular}
\label{tab:ground_stations_pt2}
\vspace{0.5em} \\
{\raggedright \centering{$\phantom{-}^{\star}$ Site list locations inferred from publicly available data} \par}
{\raggedright \centering{$\phantom{-}^{\dagger}$ Site list provided by GSaaS provider} \par}
\end{table*}

%% file: sections/appendix_contact_assumption.tex
\section{Surrogate Optimization Assumption Analysis}
\label{sec:appendix_prediction_assumption}

This paper employs surrogate optimization approach where the ground station network is optimized over a short simulation window that is less than the entire mission duration. This is assumed to be a good proxy for optimizing over the entire mission and provides the benefit of reducing the problem size, thereby improving runtime performance. It also serves to address the challenge of accurately predicting contact windows over a potentially multi-year space missions by eliminating the need to simulate the entire mission duration. The surrogate optimization is appropriate so long as the distribution of contacts and contact properties of the surrogate problem is similar to that of longer propagations.

\begin{figure*}[ht!]
  \centering
  % Subfigure 1
  \begin{subfigure}{0.3\textwidth}
    \centering
    \includegraphics[width=\textwidth]{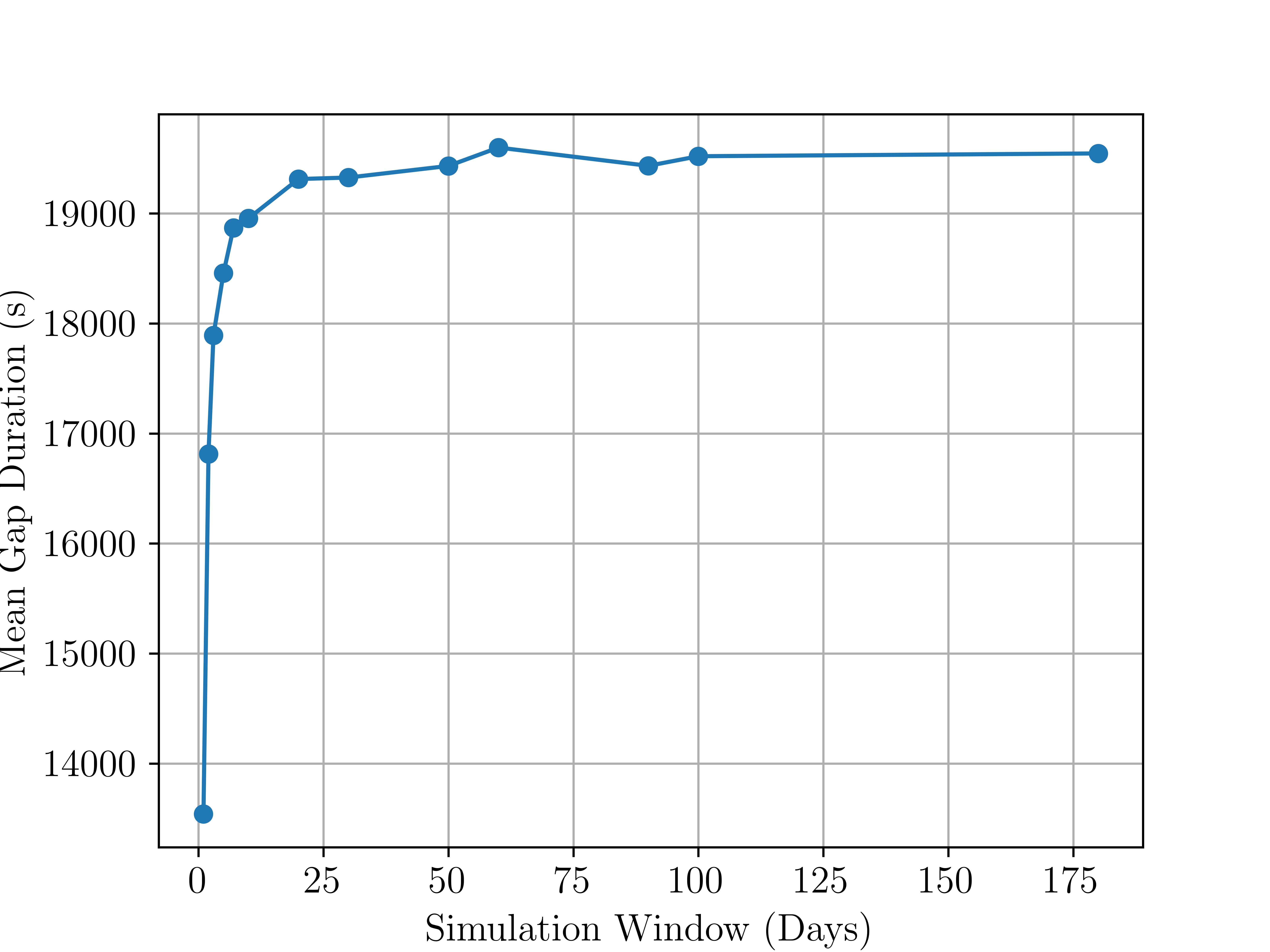}%    \caption{First Subfigure}
  \label{fig:contact_assumption_mean_gapduration}
  \end{subfigure}
  \hfill
  % Subfigure 2
  \begin{subfigure}{0.3\textwidth}
    \centering
    \includegraphics[width=\textwidth]{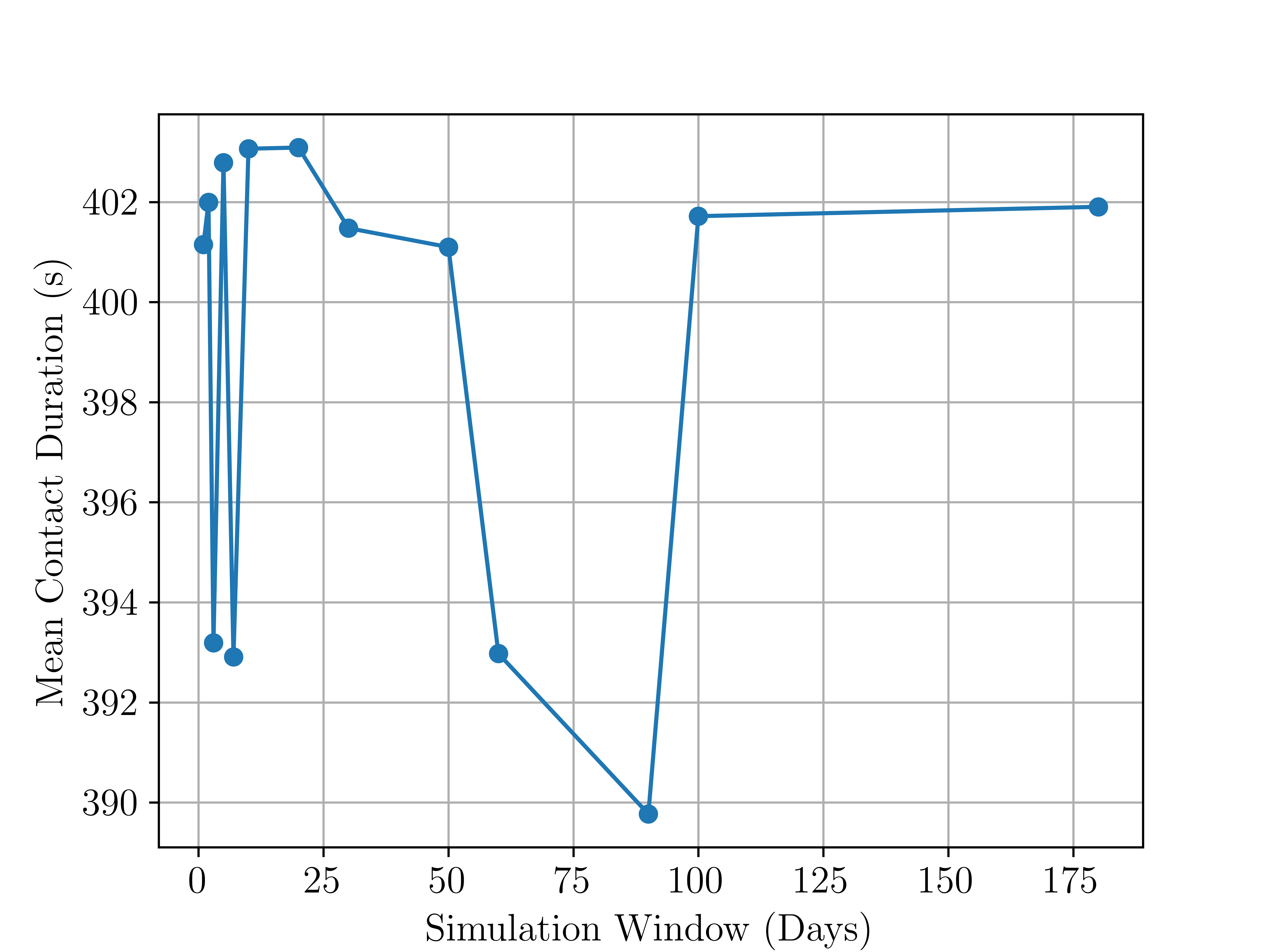}%    \caption{Second Subfigure}
    \label{fig:contact_assumption_mean_contactduration}
  \end{subfigure}
  \hfill
  % Subfigure 3
  \begin{subfigure}{0.3\textwidth}
    \centering
    \includegraphics[width=\textwidth]{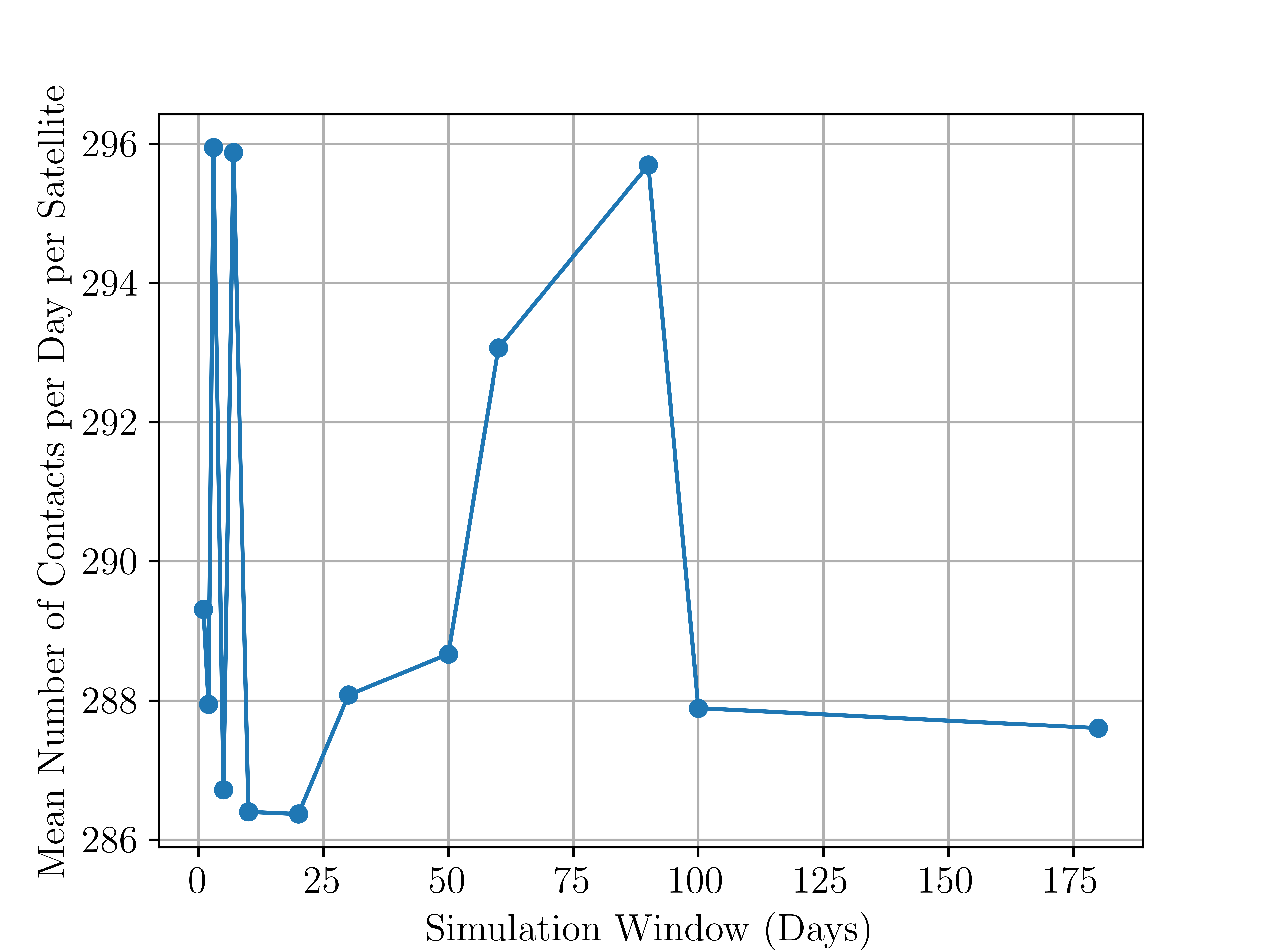}
%    \caption{Runtime}
    \label{fig:sub3}
  \end{subfigure}
  \caption{Contact simulation statistics for 100 satellites propagated over 1, 2, 3, 5, 7, 10, 20, 30, 50, 60, 90, 100, and 180-day simulation windows.  Mean contact gap duration (left), mean contact duration (middle) and mean number of contacts (right).}
  \label{fig:contact_prop_test}
\end{figure*}

To test whether this was an appropriate assumption we randomly sampled 100 different active satellites with altitudes between \SI{300}{\km} and \SI{1000}{\km}, propagated their orbits for varying time durations, computed the contact opportunities over each duration, then computed statistics on the distribution of contacts and contact properties. \Cref{fig:contact_prop_test} shows the results of this experiment. We can see that the mean gap duration starts to level out after 7 days of propagation and is near the steady-state value by 20 days of propagation. While there appears to be significant variation in the mean contact duration and average number of contacts per day, in absolute numbers the total range of variation is relatively small at 13 seconds and 10 contacts per day respectively. The cause for this variation needs to be further investigated, though may be artifacts of the propagation method or the sample size may not be large enough. From these results we recommend a simulation optimization window of at least 7 days, however 20 days is preferred when computationally possible.

%\begin{figure*}[h!t]
%\centering
%\includegraphics[width=0.3\textwidth]{figures/simwindow_appendix/mean_gap_duration.png}
%\caption{Mean Gap Duration for 100 satellites propagated for 1, 2, 3, 5, 7, 10, 20, 30, 50, 60, 90, 100, and 180-day simulation windows. The mean reaches steady state after the 7-day simulation window.} 
%\label{fig:contact_assumption_mean_gapduration}
%\end{figure*}
%
%\begin{figure*}[h!t]
%\centering
%\includegraphics[width=0.3\textwidth]{figures/simwindow_appendix/mean_contact_duration.png}
%\caption{Mean Contact Duration for 100 satellites propagated for 1, 2, 3, 5, 7, 10, 20, 30, 50, 60, 90, 100, and 180-day simulation windows.} 
%\label{fig:contact_assumption_mean_contactduration}
%\end{figure*}
%
%\begin{figure*}[h!t]
%\centering
%\includegraphics[width=0.3\textwidth]{figures/simwindow_appendix/mean_contacts_per_day_per_satellite.png}
%\caption{Mean Number of Contacts Per Day Per Satellite for 100 satellites propagated for 1, 2, 3, 5, 7, 10, 20, 30, 50, 60, 90, 100, and 180-day simulation windows.} 
%\label{fig:contact_assumption_mean_contacts}
%\end{figure*}

%% file: sections/appendix_min_cost_opt.tex
\section{Mission Cost Minimization Problem}
\label{sec:appendix_min_cost_opt}

The mission cost minimization problem detailed in \Cref{eqn:prob_min_cost} aims to reduce the overall expenses of the ground station network throughout the mission’s duration while ensuring that each satellite achieves a specified minimum data downlink requirement. This formulation incorporates constraints for station contact exclusion and satellite contact exclusion to maintain schedules that are physically realistic. Additionally, a constraint on the minimum contact duration is included to guarantee that each contact meets a required minimum length. The enforcement of a minimum data downlink per satellite is necessary to prevent degenerate solutions, where no ground stations or providers are selected, resulting in an optimal but non-functional zero-cost outcome. The design constants of this problem are $D^s_{min}$, the required per-satellite, minimum data downlink volume, $T^{period}$, the duration over which the minimum data downlink amount must be achieve, and $t_{step}$ the discrete increment over which the adherence to the minimum data downlink constraint is checked. The full optimization problem is
\begin{equation*}
\begin{aligned}
   	& \underset{\mathbf{c,l,p,v}}{\text{minimize}} && \sum_{(p,P) \in \mathcal{P}}e^P_{integ}p \, + \, \sum_{P \in \mathcal{P}}\sum_{(l,L) \in {\mathcal{L}}^P}e^L_{setup}l \, + \\
    & && \hspace{-2em} \frac{12 \times T_{opt}}{365.25 \times 86400 \times T_{sim}}\sum_{P\in \mathcal{P}}\sum_{(l,L) \in {\mathcal{L}}^P}e^L_{monthly}l \, + \\
    & && \sum_{S \in \mathcal{S}}\sum_{P \in \mathcal{P}}\sum_{L \in {\mathcal{L}}^P} e^{L}_{license}v^{L,S} \, + \\
    & && \hspace{-2em} \frac{T_{opt}}{T_{sim}}\sum_{P \in {\mathcal{P}}}\sum_{L \in {\mathcal{L}}^P}\sum_{(c,C) \in {\mathcal{C}}^{L}}(e^{L}_{minute}C_{duration} + e^L_{pass})c \\
    &\text{subject to} && \sum_{(c, C) \in {\mathcal{C}}^L}c = l, \; \forall \; (l,L) \in {\mathcal{L}} \\
    & && \sum_{(l, L) \in {\mathcal{L}}^P}l = p, \; \forall \; (p, P) \in {\mathcal{P}} \\
\end{aligned}
%\label{eqn:prob_min_cost}
\end{equation*}
\begin{equation*}
\begin{aligned}
    & && \sum_{c \in {\mathcal{C}}^{S,L}}c = v^{S,L}, \; \forall \; l \in {\mathcal{L}}, s \in {\mathcal{S}} \\
	& && c_i + c_j \leq 1 \; \\
   	& && \hspace{2em} \forall \; L \in {\mathcal{L}}, i,j \in \{0,\ldots,|{\mathcal{C}}^L|\}, j > i \; \text{s.t.} \\
   	& && \hspace{2em} C_{start,i} \leq C_{end,j} \\
   	& && \hspace{2em} C_{start,j} \leq C_{end,j} \\
	& && c_i + c_j \leq 1 \; \\
	& && \hspace{2em} \forall \; S \in {\mathcal{S}}, i,j \in \{0,\ldots,|{\mathcal{C}}^S|\}, j > i \; \text{s.t.} \\
	& && \hspace{2em} C_{start,i} \leq C_{end,j} \\
	& && \hspace{2em} C_{start,j} \leq C_{end,j} \\
   	& &&  x = 0 \; \text{if} \; C_{end} - C_{start} < t_{min}, \forall \; (c,C) \in {\mathcal{C}} \\
   	& && \sum_{(c,C) \in {\mathcal{C}}^S}\min(L_{dr},S_{dr})C_{duration}c \geq D^S_{min}, \\
   	& && \hspace{2em} \forall \; S \; \in {\mathcal{S}}, C \; \text{s.t.} \\
   	& && \hspace{2em} C_{start} \leq t^{window}_{end} \\
   	& && \hspace{2em} C_{end} \geq t^{window}_{start} \\
   	& && \hspace{2em} t^{window}_{end} \leftarrow t^{window}_{start} + T^{period} \\
   	& && \hspace{-1em} t^{window}_{start} \in \{t^s_{sim},t^s_{sim}+t_{step},\ldots,t^e_{sim}-T^{period}\} \\
   	& && c \in \{0, 1\} \; \forall \; c \in {\mathcal{C}} \\
   	& && p \in \{0, 1\} \; \forall \; p \in {\mathcal{P}} \\
   	& && l \in \{0, 1\} \; \forall \; l \in {\mathcal{L}} \\
   	& && v^{L,S} \in \{0, 1\} \; \forall \; s \in {\mathcal{S}}, l \in {\mathcal{L}}
\end{aligned}
\label{eqn:prob_min_cost}
\end{equation*}

%% file: sections/appendix_max_data_opt.tex
\section{Data Volume Maximization Problem}
\label{sec:appendix_max_data_opt}

The problem presented in \Cref{eqn:prob_max_data} is designed to maximize the total volume of data downlinked throughout the mission period while imposing a constraint on the maximum allowable operational costs. This formulation includes station contact exclusion and satellite contact exclusion constraints to ensure that the scheduling remains physically feasible. Additionally, a constraint on the minimum contact duration is incorporated to ensure that each contact meets the necessary minimum time requirement. The maximum operational cost constraint is needed to prevent the optimizer from selecting all available providers and locations, which would maximize downlink opportunities but result in an impractical degenerate solution. The design parameters for this problem are $E_{max}$, which denotes the highest permissible monthly operational cost, and $t_{min}$, representing the minimum acceptable duration for each contact. The full problem is
\begin{equation*}
\begin{aligned}
   	& \underset{\mathbf{x,l,p,s}}{\text{maximize}} && \frac{T_{opt}}{T_{sim}}\sum_{P \in {\mathcal{P}}}\sum_{L \in {\mathcal{L}}^P}\sum_{(c,C) \in {\mathcal{C}}^L}C_{dr}C_{duration}c \\
   	&\text{subject to} &&\sum_{(c, C) \in {\mathcal{C}}^L}c = l, \; \forall \; (l,L) \in {\mathcal{L}} \\
    & && \sum_{(l, L) \in {\mathcal{L}}^P}l = p, \; \forall \; (p, P) \in {\mathcal{P}} \\
    & && \sum_{c \in {\mathcal{C}}^{S,L}}c = v^{S,L}, \; \forall \; l \in {\mathcal{L}}, s \in {\mathcal{S}} \\
    & && c_i + c_j \leq 1 \; \\
   	& && \hspace{2em} \forall \; L \in {\mathcal{L}}, i,j \in \{0,\ldots,|{\mathcal{C}}^L|\}, j > i \; \text{s.t.} \\
\end{aligned}
%\label{eqn:prob_max_data}
\end{equation*}
\begin{equation*}
\begin{aligned}
   	& && \hspace{2em} C_{start,i} \leq C_{end,j} \\
   	& && \hspace{2em} C_{start,j} \leq C_{end,j} \\
	& && c_i + c_j \leq 1 \; \\
	& && \hspace{2em} \forall \; S \in {\mathcal{S}}, i,j \in \{0,\ldots,|{\mathcal{C}}^S|\}, j > i \; \text{s.t.} \\
	& && \hspace{2em} C_{start,i} \leq C_{end,j} \\
	& && \hspace{2em} C_{start,j} \leq C_{end,j} \\
   	& &&  x = 0 \; \text{if} \; C_{end} - C_{start} < t_{min}, \forall \; (c,C) \in {\mathcal{C}} \\
   	& && \hspace{-3em} \frac{365.25 \times 86400}{12T_{sim}}\sum_{L \in {\mathcal{L}}}\sum_{(c,C) \in {\mathcal{C}}^L}(e^L_{pass}C_{pass} + C_{pass})c \; + \\
    & && \hspace{2em} \sum_{(l,L) \in {\mathcal{L}}}e^L_{monthly}l \, \leq E_{max} \\
	& && c \in \{0, 1\} \; \forall \; c \in {\mathcal{C}} \\
	& && p \in \{0, 1\} \; \forall \; p \in {\mathcal{P}} \\
	& && l \in \{0, 1\} \; \forall \; l \in {\mathcal{L}} \\
\end{aligned}
\label{eqn:prob_max_data}
\end{equation*}

%% file: sections/appendix_max_gap_opt.tex
\section{Maximum Communications Gap Minimization Problem}
\label{sec:appendix_max_gap_opt}

The problem of minimizing the maximum communications gap is the final problem variation. This problem variation represents the design optimization study of a performance-oriented mission design that seeks to minimize communications gaps subject to station exclusion, satellite exclusion, and minimum contact duration constraints. A minimum satellite data downlink constraint also must be included otherwise the problem admits a degenerate solution where no contacts are scheduled. Additionally, a maximum monthly cost constraint needs to be introduced to prevent the degenerate solution of scheduling all contacts. The design constants of this problem are $C^{max}$, the maximum monthly cost, $D^s_{min}$, the per-satellite, minimum data downlink amount, $T^{period}$, the duration over which the minimum data downlink amount must be achieve, and $t_{step}$ the discrete increment over which the adherence to the minimum data downlink constraint is checked. The full problem is
\begin{equation*}
\begin{aligned}
    & \underset{\mathbf{x}}{\text{minimize}} && G_{max} \\ 
    & \text{subject to} && \sum_{j>i}y_{ij} = c_i, \forall \; (c,C) \in {\mathcal{C}}^S, S \in {\mathcal{S}},  \\ % \forall i \in |X^{p,l}|
    & && \hspace{2em} i,j \in \{0,\ldots,|{\mathcal{C}}^S|\}\; \text{s.t.} \\
    & && \; C_{start,j} > C_{start,i} \\
    & && (C_{start,j} - C_{end,i})y_{ij} \leq G_{max},  \\
    & && y_{ij} \leq c_i, \\
    & && y_{ij} \leq c_j, \\
    & && \sum_{(c, C) \in {\mathcal{C}}^L}c = l, \; \forall \; (l,L) \in {\mathcal{L}} \\
    & && \sum_{(l, L) \in {\mathcal{L}}^P}l = p, \; \forall \; (p, P) \in {\mathcal{P}} \\
    & && \sum_{c \in {\mathcal{C}}^{S,L}}c = v^{S,L}, \; \forall \; l \in {\mathcal{L}}, s \in {\mathcal{S}} \\
    & && c_i + c_j \leq 1 \; \\
   	& && \hspace{2em} \forall \; L \in {\mathcal{L}}, i,j \in \{0,\ldots,|{\mathcal{C}}^L|\}, j > i \; \text{s.t.} \\
   	& && \hspace{2em} C_{start,i} \leq C_{end,j} \\
   	& && \hspace{2em} C_{start,j} \leq C_{end,j} \\
	& && c_i + c_j \leq 1 \; \\
	& && \hspace{2em} \forall \; S \in {\mathcal{S}}, i,j \in \{0,\ldots,|{\mathcal{C}}^S|\}, j > i \; \text{s.t.} \\
	& && \hspace{2em} C_{start,i} \leq C_{end,j} \\
	& && \hspace{2em} C_{start,j} \leq C_{end,j} \\
   	& &&  x = 0 \; \text{if} \; C_{end} - C_{start} < t_{min}, \forall \; (c,C) \in {\mathcal{C}} \\
   	& && \sum_{(c,C) \in {\mathcal{C}}^S}C_{dr}C_{duration}c \geq D^S_{min}, \\
   	& && \hspace{2em} \forall \; S \; \in {\mathcal{S}}, C \; \text{s.t.} \\
\end{aligned}
\label{eqn:prob_min_max_gap}
\end{equation*}
\begin{equation*}
\begin{aligned}
   	& && \hspace{2em} C_{start} \leq t^{window}_{end} \\
   	& && \hspace{2em} C_{end} \geq t^{window}_{start} \\
   	& && \hspace{2em} t^{window}_{end} \leftarrow t^{window}_{start} + T^{period} \\
   	& && \hspace{0em} t^{window}_{start} \in \{t^s_{sim},t^s_{sim}+t_{step},\ldots,t^e_{sim}-T^{period}\} \\
    & && \hspace{-3em} \frac{365.25 \times 86400}{12T_{sim}}\sum_{L \in {\mathcal{L}}}\sum_{(c,C) \in {\mathcal{C}}^L}(e^L_{pass}C_{pass} + C_{pass})c \; + \\
    & && \hspace{2em} \sum_{(l,L) \in {\mathcal{L}}}e^L_{monthly}l \, \leq E_{max} \\
	& && c \in \{0, 1\} \; \forall \; c \in {\mathcal{C}} \\
	& && p \in \{0, 1\} \; \forall \; p \in {\mathcal{P}} \\
	& && l \in \{0, 1\} \; \forall \; l \in {\mathcal{L}} \\
   	& && y_{ij} \in \{0,1\} \\
\end{aligned}
\label{eqn:prob_min_max_gap}
\end{equation*}